\documentclass[aps,pra,twocolumn,superscriptaddress,a4paper,10pt]{revtex4-2}
\usepackage{amsmath,amssymb,amsthm}
\allowdisplaybreaks
\usepackage{bm}
\usepackage{here}
\usepackage{mathtools}
\usepackage{physics}
\usepackage{graphicx}
\usepackage[x11names,dvipsnames]{xcolor}
\usepackage[colorlinks=true,citecolor=blue,linkcolor=blue,urlcolor=blue]{hyperref}

\begin{document} 
\title{Measurement-Based Quantum Computation Using the Spin-1 $XXZ$ Model with Uniaxial Anisotropy} 
\author{Hiroki Ohta}
\affiliation{Department of Physics, Chuo University, Bunkyo, Tokyo 112-8551, Japan} 
\author{Aaron Merlin M\"uller}
\affiliation{Department of Materials, ETH Zurich, Vladimir-Prelog-Weg 4, Zurich,
8093, Switzerland.} 
\author{Shunji Tsuchiya}
\affiliation{Department of Physics, Chuo University, Bunkyo, Tokyo 112-8551, Japan} 

\begin{abstract}
We demonstrate that the ground state of a spin-1 $XXZ$ chain with uniaxial anisotropies, single-ion anisotropy $D$ and Ising-like anisotropy $J$, within the Haldane phase can serve as a resource state for measurement-based quantum computation implementing single-qubit gates. The gate fidelity of both elementary rotation gates and general single-qubit unitary gates composed of rotations about the $x$, $y$, and $z$ axes is evaluated, and is found to exceed 0.99 when $D$ or $J$ is appropriately tuned. Furthermore, we derive an analytic expression for the rotation-gate fidelity under the assumption that the state lies within the $\mathbb Z_2\times \mathbb Z_2$-protected Haldane phase, showing that it is determined by the postmeasurement spin-spin correlation function and the failure probability. The observed enhancement of gate fidelity in the spin-1 $XXZ$ chain originates from the strengthening of antiferromagnetic (AFM) correlations near the AFM phase, which effectively suppresses failure states.
\end{abstract}
\maketitle 
\section{INTRODUCTION}
\label{sec:I}
Quantum computation offers fundamentally new approaches to information processing by exploiting quantum mechanical features—such as superposition and entanglement—that are absent in classical computation \cite{Deutsch1989,Nielsen2011}. In recent years, measurement-based quantum computation (MBQC) \cite{Raussendorf2001,Brennen2008,Miyake2010} has been extensively studied as an alternative to the traditional quantum circuit model \cite{Deutsch1989,Barenco1995,Nielsen2011}. In MBQC, computation is carried out through adaptive single-particle measurements on an entangled many-body resource state. Once the resource state is prepared, the entire computation proceeds using only local measurements and classical processing. Although MBQC is mathematically equivalent to the circuit model \cite{Raussendorf2001}, its explicit separation between classical processing and quantum operations provides a particularly advantageous framework for quantum computation.

Well-known examples of MBQC resource states include the cluster state \cite{Briegel2001} and the Affleck-Kennedy-Lieb-Tasaki (AKLT) state \cite{Affleck1987,Affleck1988}, both of which support universal quantum computation \cite{Raussendorf2001,Raussendor2003,Brennen2008,Wei2011} and belong to symmetry-protected topological (SPT) phases \cite{Pollman2010,Chen2013,Wen2017}. An SPT phase is a class of quantum many-body states characterized by discrete symmetries and short-range entanglement \cite{Chen2010,Chen2013}. They exhibit nontrivial edge modes protected by the corresponding symmetries \cite{Hagiwara1990,Kennedy1990,Glarum1991}.
The computational power of MBQC is believed to be fundamentally rooted in the SPT order of the underlying resource state \cite{Verstraete2006,Stephen2009,Barlett2010,Miyake2010,Dominic2012,Miller2015,Raussendorf2017,Stephen2017,Raussendorf2023,yang2024}. 

The Haldane phase \cite{Haldane1983,HALDANE1983464} is a prototypical SPT phase that includes both the AKLT state and the ground state of the spin-1 antiferromagnetic (AFM) Heisenberg model \cite{Affleck1987,Kennedy1992,White1993,Pollman2010,Pollman2012}. It is characterized by a nonlocal string order parameter, reflecting hidden $\mathbb{Z}_2 \times \mathbb{Z}_2$ symmetry breaking \cite{Tasaki1991,Nijs1989,Pollman2010,Perez2008}. The computational power of the SPT phase protected by $\mathbb Z_2\times\mathbb Z_2$  symmetry has been investigated \cite{Raussendorf2023,masui2024,Dominic2012,yang2024,Raussendorf2017,Miller2015,Verstraete2006,Barlett2010,Stephen2009,Stephen2017}. In particular, it has been shown that a perfect implementation of the identity gate is possible for any state in this phase \cite{Dominic2012}, and more generally it has been proven that nonvanishing string order implies the realizability of arbitrary single-qubit rotations in a one-dimensional spin-1/2 system \cite{Raussendorf2023}. However, these results do not provide how the computational capability depends on Hamiltonian parameters and anisotropy in a given physical model, which are essential for the experimental realization of MBQC on a spin chain.

The spin-1 $XXZ$ model is a deformation of the spin-1 Heisenberg model with Ising-like anisotropy \cite{Kennedy1992,White1993,Chen2003,Pollman2010}. It has attracted much interest in condensed matter physics because of its rich ground state properties, which arises due to an interplay between anisotropy, strong correlation, and quantum fluctuation. In fact, various quantum phases emerge in the ground state of this model including the AFM and ferromagnetic (FM) phases \cite{Chen2003}.

The Haldane phase protected by the $\mathbb Z_2\times \mathbb Z_2$ symmetry emerges in the ground state of the spin-1 $XXZ$ model with single-ion anisotropy \cite{Pollman2010}. Since this phase appears in the region where the Ising-like anisotropy and the single-ion anisotropy are nonzero, it is of particular interest to investigate its computational capability in order to clarify how anisotropy affects the computational power. Previous studies have mainly focused on isotropic models, such as the AKLT model and the AFM Heisenberg model \cite{Miyake2010,Barlett2010}, while the role of anisotropy in determining the computational power has remained largely unexplored.

In this work, we numerically demonstrate that in the one-dimensional (1D) spin-1 $XXZ$ model, there exists a parameter regime in which high-fidelity quantum computation can be achieved by appropriately tuning the anisotropy. Furthermore, we find that high-fidelity single-qubit universal quantum computation can be realized by partitioning the spin-1 chain and introducing anisotropy in each block. 
To investigate the quantum computational power, we analyze the gate fidelity and derive a general expression for it assuming only that the resource state lies within the Haldane phase protected by $\mathbb{Z}_2 \times \mathbb{Z}_2$ symmetry. Consequently, the derived formula applies to any state within this phase. Under this assumption, the $z$-rotation gate fidelity depends on the postmeasurement spin correlations as 
\[
F_{R_z(\theta)}=1-\frac{\sin^2{\theta}}{2}{(1+g_{\rm corr})}-\frac{(1-\cos{\theta})}{2}g_{\rm fail}.
\]
Here, $g_{\rm corr}$ represents the postmeasurement spin-spin correlation function averaged over all measurement outcomes, and $g_{\rm fail}$ denotes the probability that the measurement fails at all sites. 

We note the relation of Ref.~\cite{Raussendorf2023} to our work. Reference~\cite{Raussendorf2023} proves, in a general and mathematical framework, that 1D abelian $G$-symmetric phases with a nonvanishing string order parameter can support universal single-qubit MBQC. Theorem 1 in Ref.~\cite{Raussendorf2023} is formulated for short-range-entangled states of a spin-1/2 chain. In contrast, our work considers a spin-1 chain. We explicitly demonstrate, through numerical analysis, that single-qubit rotation gates, as well as general single-qubit unitary gates, can be implemented with gate fidelity close to unity in the ground state of the spin-1 $XXZ$ chain near the antiferromagnetic phase boundary. Our results suggest that single-qubit gates can be experimentally implemented in spin-1 chains with uniaxial anisotropies in cold-atom systems, as proposed in Refs.~\cite{Mogerle2025,Brechtelsbauer2025}. Furthermore, the surprising aspect of our results is that we identify a physical origin of the computational power that has not been explicitly discussed before, namely that the gate fidelity within the Haldane phase is quantitatively linked to postmeasurement antiferromagnetic spin-spin correlations.

The structure of this paper is as follows. In Sec.~\ref{sec:II}, we review the AKLT model and MBQC using the AKLT state as a resource. In Sec.~\ref{sec:III}, we introduce the gate fidelity as a measure of quantum computational power and evaluate it for the AKLT state and for the ground state of the bilinear-biquadratic (BLBQ) model. In Sec.~\ref{sec:IV}, we calculate the gate fidelity of the $z$-rotation gate for the ground state of the 1D spin-1 $XXZ$ model, demonstrating that it can be enhanced by appropriately tuning the Ising-like or single-ion anisotropy. In Sec.~\ref{sec:V}, we show that arbitrary single-qubit gates can be realized by partitioning the spin-1 chain and applying anisotropy in different directions within each block. Finally, in Sec.~\ref{sec:VI}, we summarize our results and discuss future prospects.
\section{MBQC based on the AKLT state}
\label{sec:II}
In this section, we review how single-qubit gates can be implemented in MBQC using the 1D AKLT state as a resource state \cite{Affleck1987,Affleck1988,Brennen2008}.
\begin{figure}[t]
\begin{center}
\includegraphics[width=\columnwidth]{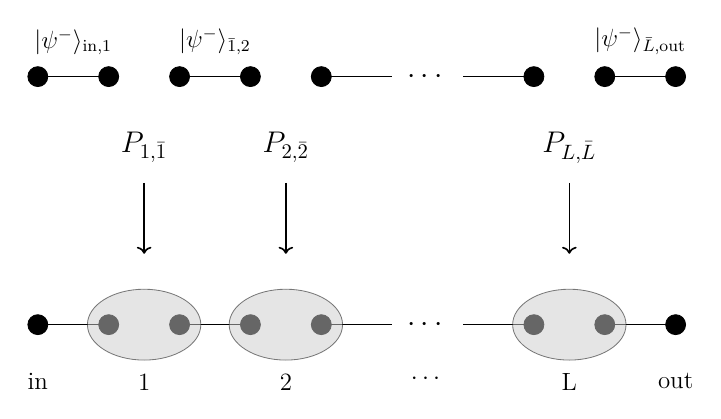}
\caption{Scematic illustration of the AKLT state. The pair of black dots connected by a line represents a spin-singlet pair of spin-1/2's, and the gray circles represent the spin-1 sites. The AKLT state is constructed by projecting spin-1/2 singlet pairs onto the physical spin-1 space. The two spin-1/2's at the ends enable input and output of single-qubit states in MBQC. }  
\label{fig:AKLT_schematic}
\end{center}
\end{figure}
\subsection{AKLT state}
\label{subsec:IIA}
We consider the system of a 1D chain of spin-1's with spin-1/2's attached at the left and right ends. The spin-1/2's at the ends enable input and output of single-qubit states in MBQC. The AKLT state is made of singlet pairs of virtual spin-1/2's, where adjacent two spin-1/2's of each neighboring spin-singlet pairs are projected onto the physical spin-1 subspace, as schematically illustrated in Fig.~\ref{fig:AKLT_schematic} \cite{Affleck1987,Affleck1988}. The wave function of the resulting state is given by
\begin{align}
\ket{{\rm AKLT}}&=
\left(\prod_{i=1}^{L} P_{i,\bar{i}} \right)\notag\\*
&\times\left[
\ket{\psi^-}_{{\rm in},1} 
\left(\prod_{i=1}^{L-1} \ket{\psi^-}_{\bar{i},i+1}\right) 
\ket{\psi^-}_{\bar{L},{\rm out}}
\right],
\label{eq:AKLT_state}
\end{align}
where $L$ is the number of spin-1's and the labels ``in'' and ``out'' denote the left and right ends, respectively. $\ket{\psi^-}$ denotes a singlet pair of two spin-1/2's,
\begin{equation}
\ket{\psi^-}=\frac{\ket{01}-\ket{10}}{\sqrt{2}}.
\label{eq:singlet_pair}
\end{equation}
Here, $\ket{0}$ and $\ket{1}$ denote the eigenstates of the Pauli matrix $Z$ corresponding to the eigenvalues $+1$ and $-1$, respectively.
$P_{i,\bar{i}}$ is an operator that projects two virtual spin-1/2's onto the physical spin-1 subspace. It is given as
\begin{align}
P_{i,\bar{i}}=
\frac{\ket{+}_i\!\bra{00}_{i,\bar{i}}}{\sqrt{3}}+\frac{\ket{\tilde 0}_i\!(\bra{01}+\bra{10})_{i,\bar{i}}}{\sqrt{6}}+\frac{\ket{-}_i\!\bra{11}_{i,\bar{i}}}{\sqrt{3}},
\label{eq:projection_onto_spin1}
\end{align}
where $\ket{+}$, $\ket{\tilde 0}$, and $\ket{-}$ are the eigenstates of the $z$ component of the spin-1 operator $\bm S=(S^x,S^y,S^z)$ corresponding to the eigenvalues $+1$, $0$, and $-1$, respectively. 

The AKLT state $|{\rm AKLT}\rangle$ is the unique ground state of the AKLT Hamiltonian \cite{Affleck1987,Affleck1988,Kennedy1992,Brennen2008}
\begin{align}
H_{{\rm AKLT}}\notag=& \bm{s}_{{\rm in}}\cdot\bm{S}_1+\sum_{i=1}^{L-1}\left[\bm{S}_i\cdot\bm{S}_{i+1}+\frac{1}{3}(\bm{S}  _i\cdot\bm{S}_{i+1})^2\right]\notag\\*  &+\bm{S}_{L}\cdot\bm{s}_{{\rm out}},  
\label{eq:AKLT_hamiltonian}
\end{align}
where $\bm{s}=(X/2,Y/2,Z/2)$ is the spin-1/2 operator ($X$ and $Y$ are the Pauli matrices). 
\begin{figure}[t]
\begin{center}
\includegraphics[width=\columnwidth]{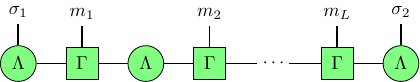}
\caption{Graphical representation of the MPS in Eq.~(\ref{eq:AKLT_MPS}) for the AKLT state. Circles and squares respectively denote MPS tensors $\Lambda$ and $\Gamma^m$ in Eq.~(\ref{eq:AKLTMPS_tensor}). Horizontal lines which connect $\Lambda$ and $\Gamma^m$ represent multiplying each matrix and vertical lines represent physical degrees of spin-$1/2$s and spin-$1$s.}  
\label{fig:AKLT_MPS}
\end{center}
\end{figure}

Equation~(\ref{eq:AKLT_state}) can be written in a matrix product state (MPS) form \cite{Fannes1992,Perez2006,SCHOLLWOCK2011,Cirac2021} , which is diagrammatically represented in Fig.~\ref{fig:AKLT_MPS}, as \cite{masui2024}
\begin{align}
\ket{{\rm AKLT}}=&\sum_{\sigma_1=0,1}\sum_{\sigma_2=0,  1}\left(\prod_{i=1}^L\sum_{m_i=+,\tilde 0,-}\right)\notag\\&\bra{\sigma_1}\Lambda\Gamma^{m_1}\Lambda\cdots\Lambda\Gamma^{m_L}\Lambda\ket{\sigma_2}\notag\\
&\times\ket{\sigma_1}_{\rm in}\ket{m_1}_1\cdots\ket{m_L}_L\ket{\sigma_2}_{\rm out}.
\label{eq:AKLT_MPS}
\end{align}
Here, MPS tensors $\Lambda$ and $\Gamma^{m}$ represent the singlet bond and the projector onto spin-1 basis, respectively, which can be written as
\begin{align}
\Lambda=\frac{XZ}{\sqrt{2}},\ \Gamma^{\pm}=\frac{I\pm Z}{\sqrt{3}},\ \Gamma^{\tilde{0}}=\sqrt{\frac{2}{3}}X.
\label{eq:AKLTMPS_tensor}
\end{align}

The AKLT state possesses a $\mathbb Z_2\times \mathbb Z_2$ symmetry corresponding to global $\pi$ rotations about the $x$, $y$, and $z$ axes \cite{Pollman2010,Pollman2012}. 
The global symmetry operators acting on the entire chain are defined as
\begin{equation}
U^{\mu}=-\sigma^{\mu}_{\rm in}\left(\prod_{i=1}^Le^{-i\pi S^{\mu}_i}\right)\sigma^{\mu}_{\rm out},
\label{eq:globalZ2Z2}
\end{equation}
where $\mu=x, y, z$, and $\sigma^\mu$  denotes the $\mu$ component of the Pauli matrix.

When $U^\mu$ acts on the AKLT state in Eq.~(\ref{eq:AKLT_MPS}), the MPS tensor transforms as \cite{Pollman2010,Chen2011,Shunch2011}
\begin{equation}
\sum_{n}(e^{-i\pi S^\mu})_{mn}\Gamma^{n}\Lambda
=(\sigma^{\mu})^{\dagger}(\Gamma^{m}\Lambda)\sigma^{\mu},
\label{eq:AKLT_MPS_tensordeform}
\end{equation}
which leads to
\begin{align}
U^\mu&|{\rm AKLT}\rangle\notag\\*
&=-\sum_{\sigma_1=0,1}\sum_{\sigma_2=0, 1}\left(\prod_{i=1}^L\sum_{m_i=+,\tilde 0,-}\right)\notag\\*
&\bra{\sigma_1}\sigma^{\mu}\Lambda(\sigma^{\mu})^{\dagger}\Gamma^{m_1}\Lambda\cdots\Lambda\Gamma^{m_L}\sigma^{\mu}(\sigma^{\mu})^{t}\Lambda\ket{\sigma_2}\notag\\*
&\times\ket{\sigma_1}_{\rm in}\ket{m_1}_1\cdots\ket{m_L}_L\ket{\sigma_2}_{\rm out}
=|{\rm AKLT}\rangle,
\label{eq:AKLT_MPS_ddeform}
\end{align}
where $t$ denotes matrix transposition.
This relation demonstrates that the AKLT state is invariant under each global $\pi$-rotation.

The same symmetry characterizes all states in the Haldane phase protected by this group, i.e., for a general state $|G\rangle$ in the Haldane phase, it satisfies 
\begin{equation}
U^\mu|G\rangle=|G\rangle,\quad \mu=x,y,z.
\label{eq:Z2Z2symmetry}
\end{equation}
We note that $U^\mu$ is the generator of the global $\mathbb{Z}_2 \times \mathbb{Z}_2$ symmetry acting on both bulk and boundary degrees of freedom.
Here, we introduce the quantity
\begin{equation}
\langle U^{\mu} \rangle=\langle G|U^{\mu}|G\rangle,\quad \mu=x,y,z,
\label{eq:expsymminout}
\end{equation}
which represents the expectation value of the global symmetry operator \cite{masui2024}.
Due to Eq.~(\ref{eq:Z2Z2symmetry}), ground states in the Haldane phase satisfy $\langle U^{\mu} \rangle= 1$. 

We stress that this quantity is conceptually distinct from the conventional string order parameter defined in Refs.~\cite{Nijs1989,Pollman2010,Perez2008}. In the literature, string operators are nonlocal bulk operators with extended but finite support that probe hidden order through bulk correlations and may take values different from unity ~\cite{Nijs1989,Pollman2010,Perez2008}. By contrast, $U^{\mu}$ is a global symmetry generator acting on both bulk and boundary degrees of freedom and commuting with symmetry-preserving Hamiltonians. Accordingly, $\langle U^{\mu}\rangle$ represents the symmetry sector of the ground state rather than a bulk string correlator.

For later convenience, we rewrite Eq.~(\ref{eq:AKLT_MPS}) in the following form \cite{Brennen2008,Miyake2010,Barlett2010}:
\begin{equation}
\begin{aligned}
&\ket{{\rm AKLT}}=
\left(\prod_{i=1}^L\sum_{m_i=+,\tilde 0,-}\right)\ket{m_1}_1\cdots\ket{m_L}_L\\
&~~~~\times\left( A^{m_L}_{\rm out}A^{m_{L-1}}_{\rm out}\dots A^{m_1}_{\rm out}X_{\rm out}Z_{\rm out}\right)\ket{\phi^+}_{\rm in,out}.
\label{eq:AKLT_MPS_simp}
\end{aligned}
\end{equation}
Here, the operator $A^m$ is defined as $A^m = \Lambda\Gamma^m$, and $|\phi^+\rangle$ denotes the Bell state
\begin{equation}
|\phi^+\rangle=\frac{|00\rangle+|11\rangle}{\sqrt{2}}.
\end{equation} 
\subsection{MBQC based on the AKLT state}
\label{subsec:IIB}
We first describe the implementation of the identity gate, which simply teleports the input state to the output spin-1/2, and then proceed to the implementation of single-qubit rotation gates. 

The identity gate can be implemented by measuring the spin-1 sites of the AKLT state in the following orthonormal basis \cite{Brennen2008,Miyake2010}:
\begin{equation}
\begin{aligned}
\ket{x}&=\dfrac{-\ket{+}+\ket{-}}{\sqrt{2}},\\
\ket{y}&=\dfrac{\ket{+}+\ket{-}}{\sqrt{2}},\\
\ket{z}&=\ket{\tilde 0}.
\label{eq:measurement_basis}
\end{aligned}
\end{equation}
\begin{figure}[t]
\begin{center}
\includegraphics[width=0.68\columnwidth]{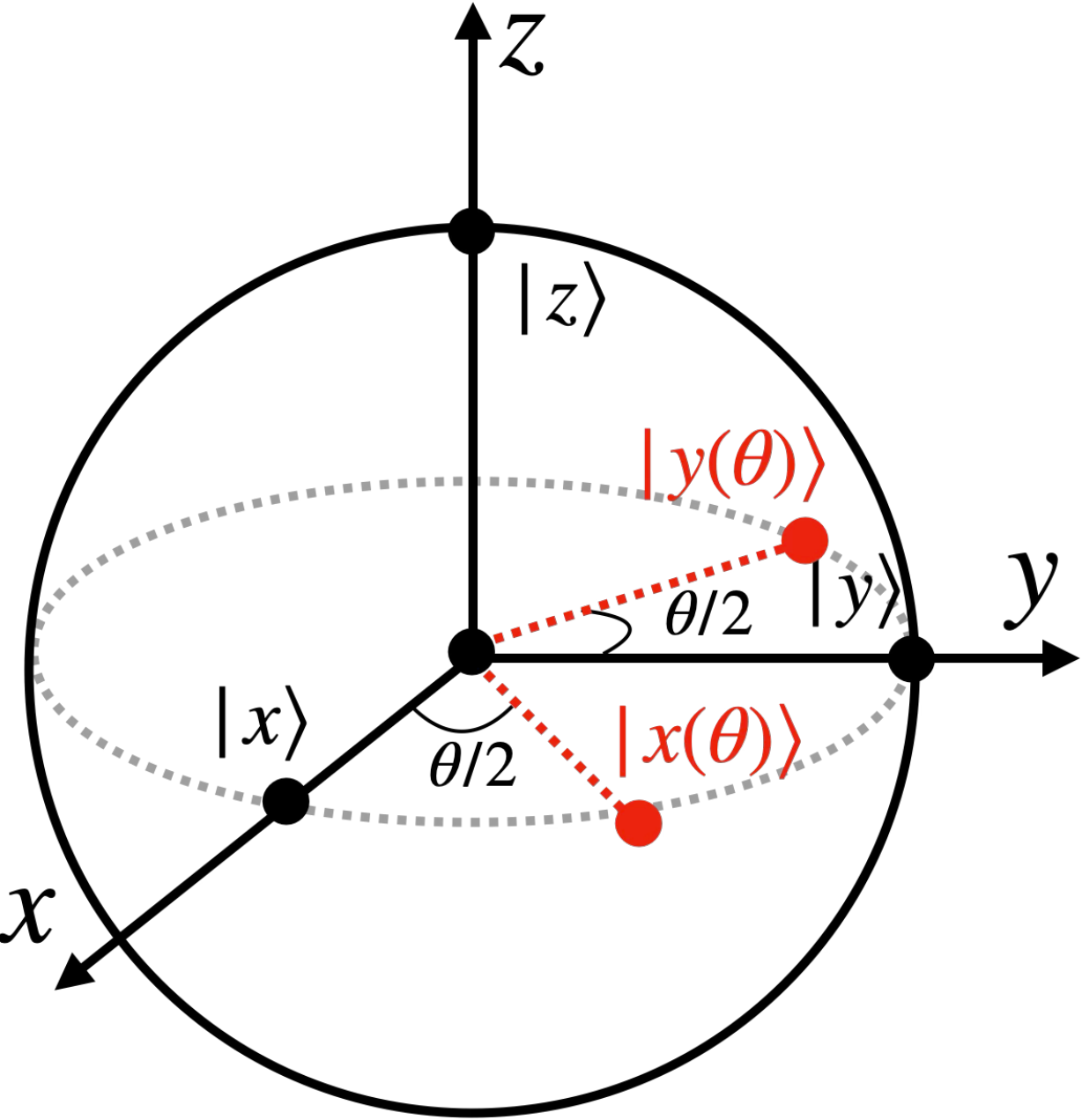}
\caption{Schematic representation of measurement basis in Eqs.~(\ref{eq:measurement_basis}) and (\ref{eq:measurement_basis_Rz}). Basis $|x(\theta)\rangle,~|y(\theta)\rangle$ are obtained by rotating basis $|x\rangle,~|y\rangle$  about $z$ axis by an angle $\theta/2$ respectively.}  
\label{fig:measurementbasis}
\end{center}
\end{figure}
Each basis state is an eigenstate of $S^x$, $S^y$, and $S^z$, respectively, with eigenvalue 0, {\rm i.e.}, $S^x\ket{x}=S^y\ket{y}=S^z\ket{z}=0$ (see Fig.~\ref{fig:measurementbasis}).
Since the basis states in Eq.~(\ref{eq:measurement_basis}) as well as the AKLT state in Eq.~(\ref{eq:AKLT_state}) are isotropic, the measurement outcomes for each site $\{\ket{x},~\ket{y},~\ket{z}\}$ are uniformly distributed, resulting in a probability of 1/3 for each basis state.

Suppose that the spin-1 at the $k$th site is measured in the basis Eq.~(\ref{eq:measurement_basis}) and the measurement outcome $\mu$ ($\mu=x,y,z$) is obtained, the state after the measurement can be written, using Eq.~(\ref{eq:AKLT_MPS_simp}), as
\begin{align}
&|\mu\rangle_k\langle{\mu}|{\rm AKLT}\rangle\notag\\*
&= \left(\prod_{i=1, i\neq k}^L\sum_{m_i=+,\tilde 0, -}\right)\ket{m_1}_1\cdots|\mu\rangle_k\cdots\ket{m_L}_L\notag\\*
&~~~\times\left[\left(  A^{m_L}_{\rm out}A^{m_{L-1}}_{\rm out}\dots A^{m_{k+1}}_{\rm out}\right)
\tilde{A}^{\mu_k}_{\rm out}\right.\notag\\*
&~~~\times\left.\left( A^{m_{k-1}}_{\rm out}\dots A^{m_1}_{\rm out}\right)
\Lambda_{\rm out}
\right] |\phi^+\rangle_{\rm in,out},
\label{eq:AKLT_aftermeasure}
\end{align}
where $\tilde A^{\mu_k }\propto\sigma^{\mu_k}~(\mu_k=x, y, z)$ is the MPS tensor after the measurement.
Here, we neglect normalization of Eq.~(\ref{eq:AKLT_aftermeasure}). Measuring all spin-1's at $1\le i\le L$, the resulting state for the input and output spin-1/2's is obtained as 
\begin{equation}
\begin{aligned}
&\left[\left(\sigma^{{\mu}_L}_{\rm out}\sigma^{{\mu}_{L-1}}_{\rm out}\dots\sigma^{{\mu}_1}_{\rm out}\right)Y_{\rm out}\right]|\phi^+\rangle_{\rm in,out},
\label{eq:state_afteroverallmeasurement_IdG}
\end{aligned}
\end{equation}
where $\{\mu_1, \mu_2, \dots, \mu_L\}$ [$\mu_i=x,y,z$ ($1\le i\le L$)] represent the measurement outcomes.
In the above equation, the Pauli matrices before $\ket{\phi^+}_{\rm in, out}$ are by-products \cite{Raussendorf2001,Raussendor2003,morimae2012} and can be removed via local operations and classical communication (LOCC) \cite{Bennet1993} based on a sequence of measurement outcomes.  
After the measurements, we obtain the ideal state $|\phi^+\rangle_{\rm in,out}$. 

Quantum teleportation of an arbitrary single-qubit state $|\chi\rangle=a|0\rangle+b|1\rangle$ ($|a|^2+|b|^2=1$) between the input and output spin-1/2's can be performed using the Bell state $|\phi^+\rangle_{\rm in,out}$ \cite{Bennet1993,Gottesman1999}.
By projecting its complex conjugate $|\chi^*\rangle$ onto the input spin-1/2, one obtains the teleported state at the output as
\begin{equation}
\langle \chi^*|_{\rm in}|\phi^+\rangle_{\rm in,out}=|\chi\rangle_{\rm out}.
\end{equation}
Since the projection of $|\chi^*\rangle$ commutes with the measurements of spin-1's and with the elimination of the by-products, this projection can be performed prior to the MBQC process as the initialization of the input state. Thus, obtaining $|\phi^+\rangle_{\rm in,out}$ after the measurements of all spin-1 sites indicates the realization of the identity gate with the AKLT state serving as a resource state.

Single-qubit rotation gates must be realized to implement arbitrary single-qubit gates, since any single-qubit unitary can be decomposed into a sequence of rotation gates.
Here, we describe the implementation of a rotation gate about the $z$-axis using the AKLT state as a resource state. For this purpose, we take the measurement basis \cite{Brennen2008,Miyake2010} 
\begin{equation}
\begin{aligned}
\ket{x(\theta)}&=\dfrac{-e^{-i\frac{\theta}{2}}\ket{+}+e^{i\frac{\theta}{2}}\ket{-}}{\sqrt{2}},\\
\ket{y(\theta)}&=\dfrac{e^{-i\frac{\theta}{2}}\ket{+}+e^{i\frac{\theta}{2}}\ket{-}}{\sqrt{2}},\\
\ket{z}&=\ket{\tilde 0}.
\label{eq:measurement_basis_Rz}
\end{aligned}
\end{equation}
These basis states are obtained by rotating the basis states in Eq.~(\ref{eq:measurement_basis}) about $z$-axis by an angle $\theta/2$ (see Fig.~\ref{fig:measurementbasis}). In fact, these basis states are the eigenstates of the rotated spin operators $S^{\mu(\theta)} = e^{-i\frac{\theta S^z}{2}} S^{\mu} e^{i\frac{\theta S^z}{2}}$ ($\mu=x,y,z$) with the corresponding eigenvalue $0$. Analogous to the identity gate, the measurement outcomes for each site are uniformly distributed, resulting in a probability of 1/3 at each basis state. 

In the case where the measurement outcome at $k$th site is $\mu(\theta)~(\mu(\theta)=x(\theta),~y(\theta),~z)$, the state after measurement can be written, using Eq.~(\ref{eq:AKLT_MPS_simp}), as
\begin{equation}
\begin{aligned}
&|\mu(\theta)\rangle_k\langle{\mu(\theta)}|{\rm AKLT}\rangle\\
=&\left(\prod_{i=1, i\neq k}^L\sum_{m_i=+,\tilde 0, -}\right)\ket{m_1}_1\cdots|\mu(\theta)\rangle_k\cdots\ket{m_L}_L\\
&\times\left[\left( \prod_{i=L}^{k+1} A^{m_i}_{\rm out} \right)
\tilde{A}^{\mu_k(\theta)}_{\rm out}
\left( \prod_{j=k-1}^{1} A^{m_j}_{\rm out} \right)
\Lambda_{\rm out}
\right] |\phi^+\rangle_{\rm in,out}.
\label{eq:AKLT_aftermeasure_Rz}
\end{aligned}
\end{equation}
\\
Here ${\tilde{A}}^{\mu_k(\theta)}$ represents the MPS tensor after the measurement, which is given by
\begin{equation}
\tilde{A}^{\mu_k(\theta)}_{\rm out}=\sum_{m_k=+,\tilde 0.-}\langle{\mu_k(\theta)}|m_k\rangle_k A^{m_k}_{\rm out}.
\end{equation}
Depending on the measurement outcome, this operator takes the form of
\begin{equation}
\begin{aligned}
\tilde{A}^{x_k(\theta)}&\propto R_z(\theta)X,\\
\tilde{A}^{y_k(\theta)}&\propto R_z(\theta)Y,\\
\tilde{A}^{z_k}&\propto Z.
\label{eq:xyz_MPS_1}
\end{aligned}
\end{equation}
Here, we neglect normalization of Eq.~(\ref{eq:AKLT_aftermeasure_Rz}). $R_z(\theta)=\exp(-{i\theta Z/2})$ denotes the spin-1/2 rotation operator about the $z$-axis by an angle $\theta$. According to Eq.~(\ref{eq:xyz_MPS_1}), the measurement outcome at each site takes either $x(\theta)$ or $y(\theta)$ with probability $2/3$, resulting in the successful application of the desired rotation gate. 
Meanwhile, the measurement outcome $z$ occurs with probability 1/3. In this case, only the by-product $Z$ is applied and the desired rotation gate fails.

We adopt the measurement protocol proposed in Ref.~\cite{masui2024} in order to implement the rotation gate with high accuracy:
\begin{enumerate}
\item[(1)] Measure the first spin-1 in the basis Eq.~(\ref{eq:measurement_basis_Rz}).
\item[(2)] If the measurement fails (i.e., the outcome is $z$), proceed to the next spin-1 and repeat the measurement until a successful outcome is obtained.
\item[(3)] Once the measurement succeeds at a certain site $k$, perform measurements on the remaining spin-1's $i = k+1, \dots, L$ using the basis for the identity gate Eq.~(\ref{eq:measurement_basis}).
\end{enumerate}
In this protocol, if the measurement outcomes at all sites are $z$, all measurements fail, and only the identity gate is applied after elimination of the by-products via LOCC. Since the probability of this event is $1/{(3^L)}$, for sufficiently large $L$, the failure probability approaches zero and  successful implementation of the rotation gate is ensured.

In the case where the measurement succeeds at the $k$th site, the state after all measurements is given by
\begin{align}
&\Bigg[\left(\prod_{i=L}^{k+1}\sigma^{{\mu}_i}_{\rm out}\right)R_{z,{\rm out}}(\theta)(X_{\rm out})^{\delta_{\nu_k(\theta), x_k(\theta)}}(Y_{\rm out})^{\delta_{\nu_k(\theta), y_k(\theta)}}\notag\\
&\qquad\qquad\qquad\times (Z_{\rm out})^{k-1}Y_{\rm out}\Bigg]|\phi^+\rangle_{\rm in,out}\notag\\
&=\Bigg[\left(\prod_{i=L}^{k+1}\sigma^{{\mu}_i}_{\rm out}\right)(X_{\rm out})^{\delta_{\nu_k(\theta), x_k(\theta)}}(Y_{\rm out})^{\delta_{\nu_k(\theta), y_k(\theta)}}\notag\\
&~~\times (Z_{\rm out})^{k-1}Y_{\rm out}\Bigg]R_{z,{\rm out}}(\theta)|\phi^+\rangle_{\rm in,out},
\label{eq:state_afteroverallmeasurement_1}
\end{align}
where $(\mu_L,\cdots,\mu_{k+1},\nu_k(\theta),z_{k-1},\cdots,z_{1})~(\mu=x,y,z,~{\rm and}~\nu=x,y)$ represent all measurement outcomes, and $\delta_{\nu_k(\theta),y_k(\theta)}$ and $\delta_{\nu_k(\theta),x_k(\theta)}$ denote Kronecker deltas.
Since the by-product operators can be removed via LOCC, from Eq.~(\ref{eq:state_afteroverallmeasurement_1}), we obtain the desired state $R_{z,\rm out}(\theta)|\phi^+\rangle_{\rm in,out}$ after the MBQC process.
By projecting $|\chi^*\rangle$ onto the input spin-1/2 of $R_{z,\rm out}(\theta)|\phi^+\rangle_{\rm in,out}$, the input state $|\chi\rangle$ is teleported to the output spin-1/2 with the rotation gate $R_z(\theta)$ applied \cite{Gottesman1999}:
\begin{equation}
\langle{\chi^*}|_{\rm in}R_{z,{\rm out}}(\theta)_{\rm out}|\phi^+\rangle_{\rm in,out}=R_{z,{\rm out}}(\theta)|{\chi}\rangle_{\rm out}.
\end{equation}
Analogous to the case of the identity gate, this projection can be carried out prior to the MBQC process.
Therefore, the AKLT state serves as a resource state for implementing the rotation gate.

The rotation gates about $x$ and $y$ axes can be realized analogously to the rotation gate about $z$ axis. When performing the $y$-rotation gate, since $y$-rotation gate $R_y(\theta)$ anticommutes with by-products, the measurement angle is reversed.
\section{Gate fidelity}
\label{sec:III}
\subsection{Definition of gate fidelity}
\label{subsec:IIIA}
We evaluate the computational power of a given many-body state as a resource for MBQC using gate fidelity \cite{Fujii2013, masui2024}, defined as
\begin{equation}
F_U=\Tr_{{\rm in,out}}\left[\rho_U|\psi_U\rangle_{\rm in, out}\langle\psi_U|\right].
\label{eq:GateFidelity}
\end{equation}
Here, $\ket{\psi_U}_{\rm in,out}=U_{\rm out}\ket{\phi^+}_{\rm in,out}$ denotes the post-MBQC state of the input and output spin-1/2's for the unitary gate $U$ on the AKLT state.
An input single-qubit state $\ket{\chi}$ is teleported, with $U$ applied, by projecting $\bra{\chi^*}$ onto the input state of $\ket{\psi_U}_{\rm in,out}$.  
The density matrix $\rho_U$ is the state after the MBQC process using the resource state $\rho$ \cite{Fujii2013, masui2024,Wang2014},
\begin{equation}
\rho_U = {\rm Tr}\left[\sum_{\bm{m}} B_{{\rm out}}^{(\bm{m})} \mathcal{P}_{\bm{m}} \rho \mathcal{P}_{\bm{m}} \left(B_{{\rm out}}^{(\bm{m})}\right)^{\dagger} \right],
\label{eq:state_aftermeasurement}
\end{equation}
where the trace is taken over all spin-1's. The index $\bm{m}$ denotes the sequence of measurement outcomes for spin-1 sites in the MBQC process,
\begin{equation}
\bm{m}=(m_1,m_2,\ldots,m_L),
\end{equation}
where $m_i$ ($1\le i\le L$) represents the measurement outcome of the $i$th spin-1 site.
$\mathcal{P}_{\bm{m}}$ and $B_{{\rm out}}^{(\bm{m})}$ denote, respectively, the projection operator corresponding to the projective measurement outcome $\bm m$ and the by-product operator acting on the output qubit that arises from the measurement $\mathcal P_{\bm m}$. 
Gate fidelity measures how close the resource state after the completion of the MBQC process $\rho_U$ is to the ideal state $\ket{\psi_U}$.
\subsection{Gate fidelity of rotation gates using the AKLT state as a resource}
\label{subsec:IIIB}
As an example, we evaluate the gate fidelity of the rotation gate $R_z(\theta)$ within the MBQC protocol described in the previous section, taking the AKLT state as a resource. The post-MBQC state is given as 
\begin{equation}
|\psi_{R_z(\theta)}\rangle_{\rm in,out}=R_{z,\rm out}(\theta)|\phi^+\rangle_{\rm in,out}. 
\end{equation}
Failure of measurements of all spin-1's occurs with probability $1/(3^L)$, resulting in only the identity gate being applied. Thus, the resource state after the MBQC protocol becomes a mixed state \cite{masui2024}, given by
\begin{align}
\rho_{R_z(\theta)}=(1-\frac{1}{3^L})|\psi_{R_z(\theta)}\rangle\langle\psi_{R_z(\theta)}|+\frac{1}{3^L}|\phi^+\rangle\langle\phi^+|.
\label{eq:AKLT_afterMBQC}
\end{align}
Plugging $\rho_{R_z(\theta)}$ and $\ket{\psi_{R_z(\theta)}}$ into Eq.~(\ref{eq:GateFidelity}), the gate fidelity is obtained as
\begin{align}
F_{R_z(\theta)}=1-\frac{1}{2\cdot3^L}(1-\cos{\theta}).
\label{eq:AKLT_RzGF}
\end{align}
Equation~(\ref{eq:AKLT_RzGF}) indicates that the gate fidelity approaches unity for large $L$ \cite{masui2024}.

In general, the gate fidelity of $R_z(\theta)$ with the resource state $|{G}\rangle$ can be written as \cite{masui2024}
\begin{align}
F_{R_z(\theta)}=&\frac{1}{4}+\frac{1}{4}\langle U^z \rangle
-\frac{\cos\theta}{4}[T_{xx}(\theta)+T_{yy}(\theta)]\notag\\
&-\frac{\sin\theta}{4}[T_{xy}(\theta)-T_{yx}(\theta)],
\label{eq:Rz_GF_expand}
\end{align}
where $U^z$ is the $z$ component of  $\mathbb{Z}_2\times\mathbb{Z}_2$ symmetry operators in Eq.~(\ref{eq:globalZ2Z2}).
$T_{\mu\nu}(\theta)$ ($\mu,\nu=x,y$) is given as
\begin{align}
T_{\mu\nu}(\theta)&=\sum_{k=1}^{L}\sum_{\bm{m}_k} \bra{G}\!{\sigma}^{\mu}_{{\rm in}} \notag\\
&\times\left[\mathcal{P}_{\bm{m}_k}\! 
\left( \prod_{i=1}^{k} e^{-i \pi {S_i^{\mu(\theta)}}} \right)\!\!
\left( \prod_{j=k+1}^{L} e^{-i \pi S^{\mu}_j} \right) \right]\!\!{{\sigma}^{\nu}_{\rm out}}\!\ket{G}\notag\\
&+\langle{G}|\sigma^{\mu}_{\rm in}\!\!\left[\left(\prod_{i=1}^L|z\rangle_i\langle z|\right)\!\!\left(\prod_{j=1}^L{e^{-i\pi S^{\mu}_j}}\right)\right]\!\!\sigma^{\nu}_{\rm out}|{G}\rangle.
\label{eq:Rz_GF_eachterm}
\end{align}
Here, $\bm{m}_k$ denotes a sequence of measurement outcomes in which the measurement first succeeds at site $k$ ($1\le k\le L$) 
\begin{equation}
\bm{m}_k=(z_1,\ldots,z_{k-1},\mu_k(\theta),\nu_{k+1},\ldots,\nu_{L}),
\end{equation}
where $\mu=x,y$ and $\nu=x,y,z$. In the cases where one of the products in the first term of Eq.~(\ref{eq:Rz_GF_eachterm}) becomes empty, namely for $k=1$ and $k=L$, it represents the identity operator.
$\mathcal{P}_{\bm{m}_k}$ denotes the projection operator associated with the measurement outcome $\bm{m}_k$:
\begin{align}
\mathcal{P}_{\bm{m}_k} =\left(\prod_{i=1}^{k-1}|z\rangle_i\langle z|\right)|{\mu}(\theta)\rangle_{k}\langle\mu(\theta)|
\left(\prod_{j=k+1}^{L}{|\nu\rangle_j\langle\nu|}\right),
\label{eq:POVM_success}
\end{align}
where $\mu=x,y$ and $\nu=x,y,z$. 
A derivation of Eq.~(\ref{eq:Rz_GF_expand}) is provided in Appendix \ref{sec:Appendix. A}. 

Setting $\theta=0$ in Eq.~(\ref{eq:Rz_GF_expand}), we obtain the gate fidelity of the identity gate \cite{masui2024},
\begin{equation}
F_I = \frac{1}{4}+\frac{1}{4}\langle U^z \rangle-\frac{1}{4}[T_{xx}(0)+T_{yy}(0)].
\label{eq:identityGF}
\end{equation}
$T_{\mu\mu}(0)$ ($\mu=x,y$) reduces as
\begin{align}
T_{\mu\mu}(0)&
=\langle{G}|\sigma^{\mu}_{\rm in}\left(\prod_{i=1}^L{e^{-i\pi S^{\mu}_i}}\right)\sigma^{\mu}_{\rm out}|{G}\rangle\notag\\*
&=-\langle U^{\mu} \rangle.
\label{eq:IdGF_eachterm}
\end{align}
Here, we use the completeness relation, 
\begin{equation}
\sum_{\bm{m}_k}\mathcal{P}_{\bm{m}_k}+\prod_{i=1}^L|z\rangle_{i}\langle{z}|=I^{\otimes L}.
\label{eq:projection_completeness}
\end{equation}
Thus, Eq.~(\ref{eq:identityGF}) reduces to a simple expression \cite{masui2024},
\begin{equation}
F_I=\frac{1}{4}+\frac{1}{4}\sum_{\mu=x,y,z}{\langle U^{\mu} \rangle} .
\end{equation}
From this expression, $F_I$ becomes unity for ground states in the Haldane phase satisfying $\langle U^\mu \rangle = 1$~\cite{masui2024}, which is associated with the emergence of edge Bell entanglement after the sequence of measurements. 
\begin{figure}[t]
\begin{center}
\includegraphics[width=\columnwidth]{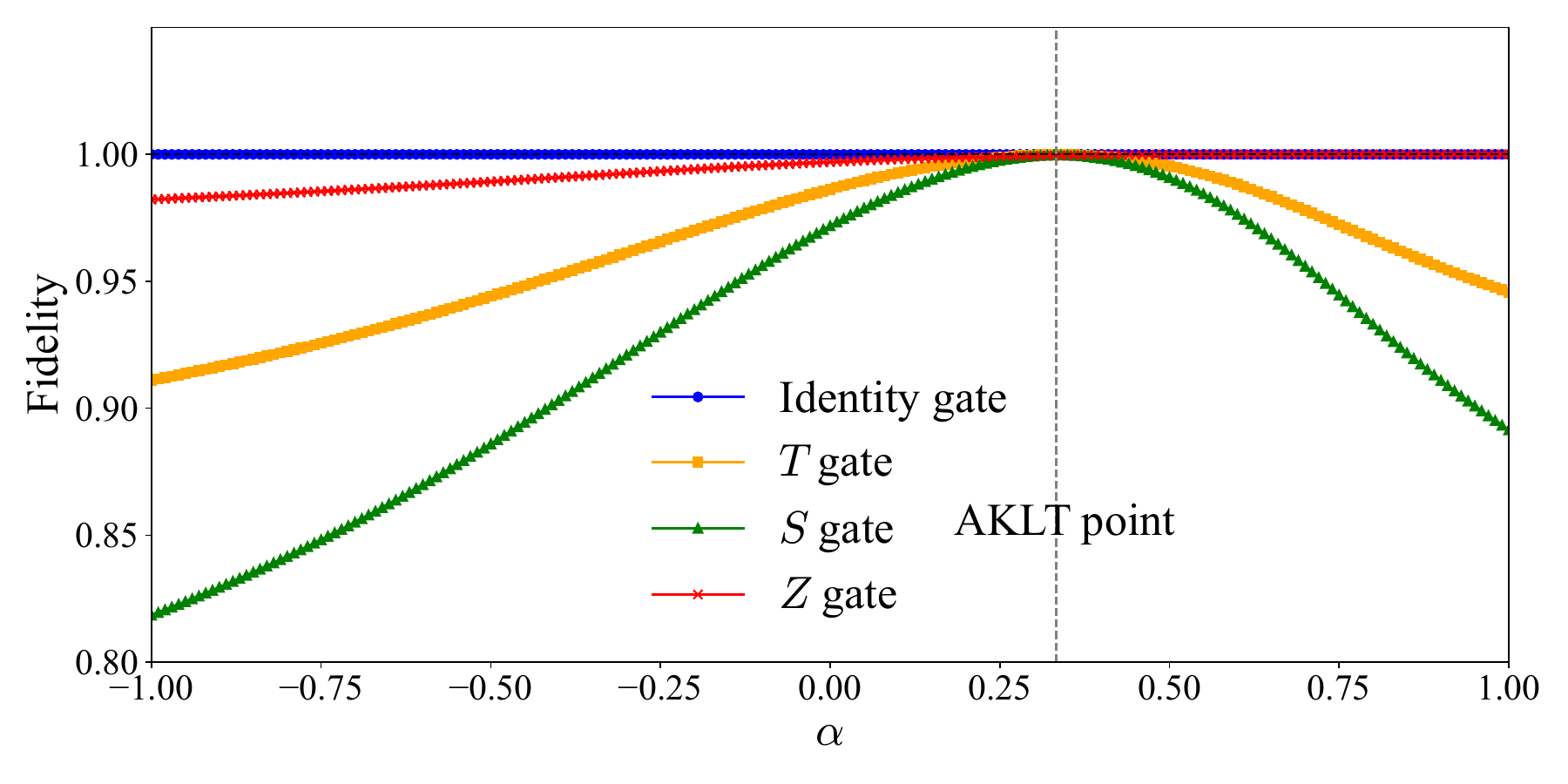}
\caption{Gate fidelity of $R_z(\theta)$ for the ground state of the BLBQ model with system size $L=7$ and $\alpha\in [-1,1]$. The gate fidelities of the identity, $T$, $S$, and $Z$ gates (corresponding to $\theta=0$, $\pi/4$, $\pi/2$, $\pi$, respectively) are calculated. The ground state is obtained using the density matrix renormalization group (DMRG) method \cite{White1992,Ostlund1995,SCHOLLWOCK2011,itensor} with a maximum bond dimension of 100. The gate fidelity is evaluated using Eqs.~(\ref{eq:Rz_GF_expand}) and (\ref{eq:Rz_GF_eachterm}).}  
\label{fig:BLBQ_GF}
\end{center}
\end{figure}
\subsection{BLBQ model}
\label{subsec:IIIC}
To demonstrate the practical use of gate fidelity in Eq.~(\ref{eq:Rz_GF_expand}), we now consider a specific family of Hamiltonians whose ground states are classified in the $\mathbb{Z}_2 \times \mathbb{Z}_2$-protected Haldane phase. Specifically, we study the BLBQ model \cite{Affleck1988,Schollw1996},
\begin{align}
H_{{\rm BLBQ}}(\alpha) &= \bm{s}_{{\rm in}}\cdot\bm{S}_1
+\sum_{i=1}^{L-1} \left[\bm{S}_i \cdot \bm{S}_{i+1} + \alpha(\bm{S}_i \cdot \bm{S}_{i+1})^2 \right]\notag\\*
&+\bm{S}_{L}\cdot\bm{s}_{{\rm out}},
\label{eq:BLBQ_hamiltonian}
\end{align}
where $\alpha$ is the parameter that controls the ratio between the bilinear and biquadratic interactions.  
The Hamiltonian in Eq.~(\ref{eq:BLBQ_hamiltonian}) becomes the AFM Heisenberg model at $\alpha=0$ and the AKLT Hamiltonian in Eq.~(\ref{eq:AKLT_hamiltonian}) at $\alpha=1/3$. 

We now focus on the ground state of Eq.~(\ref{eq:BLBQ_hamiltonian}) that belongs to the Haldane phase, and evaluate the gate fidelity.
Figure~\ref{fig:BLBQ_GF} shows the gate fidelity of $R_z(\theta)$ evaluated with respect to the ground state of Eq.~(\ref{eq:BLBQ_hamiltonian}) as a function of $\alpha$ within the Haldane phase. Here, we identify the Haldane phase as the region in which the gate fidelity of the identity gate is equal to unity. The gate fidelity of the $T$ gate [$R_z(\pi/4)$], $S$ gate [$R_z(\pi/2)$], and Pauli $Z$ gate [$R_z(\pi)$] takes unity at the AKLT point and decays continuously as $\alpha$ deviates from this point.  

Based on this result, one might expect that quantum computation becomes difficult in the states away from the AKLT state. However, as we will show in the next section, even in the Haldane phase of a different Hamiltonian that does not possess AKLT points, it is possible to implement a single-qubit gate with high-fidelity by tuning the anisotropy of the Hamiltonian.
\section{Computational Power in the Anisotropic Haldane Phase}
\label{sec:IV}
In this section, we investigate the quantum computational power of the ground state of the spin-1 $XXZ$ model, focusing on the Haldane phase in the presence of anisotropy.
\subsection{XXZ model}
\label{subsec:IVA}
We now consider the 1D spin-1 $XXZ$ model with uniaxial single-ion anisotropy \cite{Chen2003,Kennedy1992}. The Hamiltonian of this model with two spin-1/2's attached at the
\begin{figure*}
\begin{center}
\includegraphics[width=\linewidth]{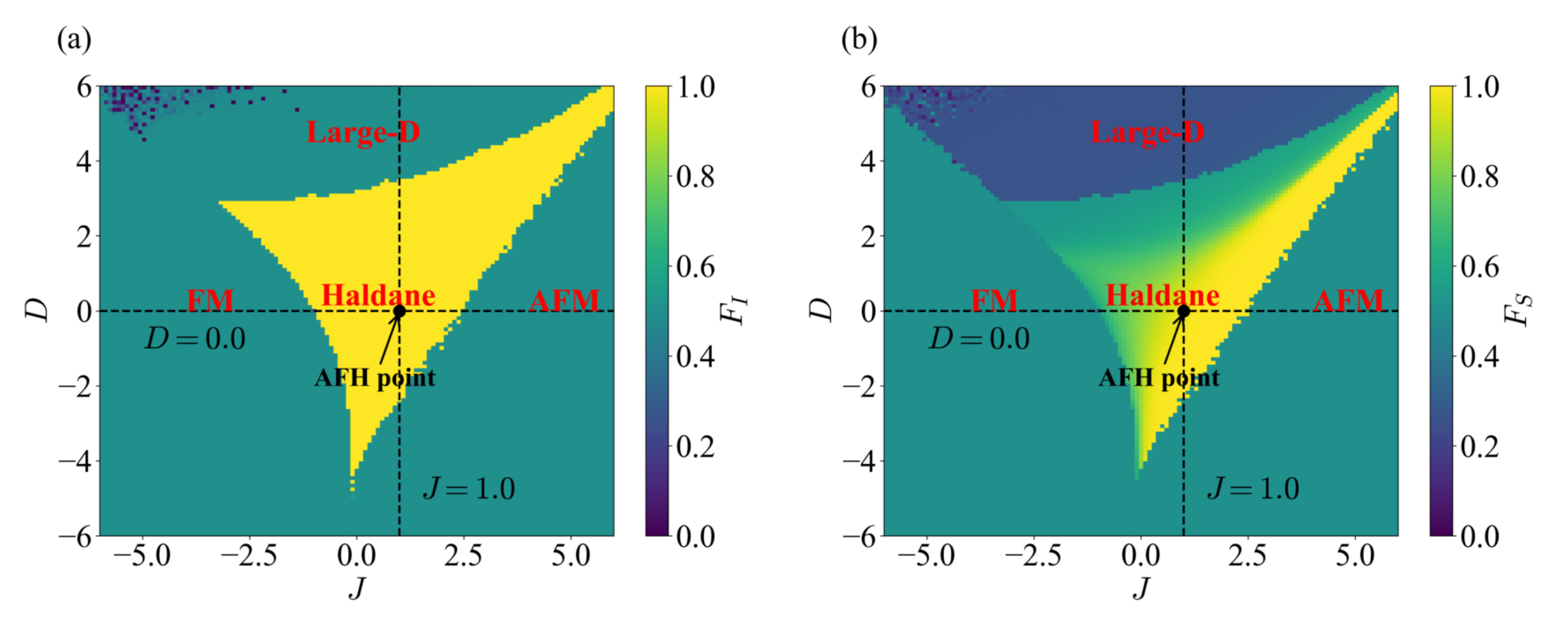}
\caption{(a) The gate fidelity of the identity gate ($\theta = 0$) and (b) the gate fidelity of the $S$ gate ($\theta = \pi/2$) for the ground state of the $XXZ$ model with the system size $L=15$ and $J,D\in[-6,6]$. The ground state is obtained using DMRG  with a maximum bond dimension of 100.}
\label{fig:XXZ_Id_S_GF}
\end{center}
\end{figure*}
\begin{figure*}
\begin{center}
\includegraphics[width=\linewidth]{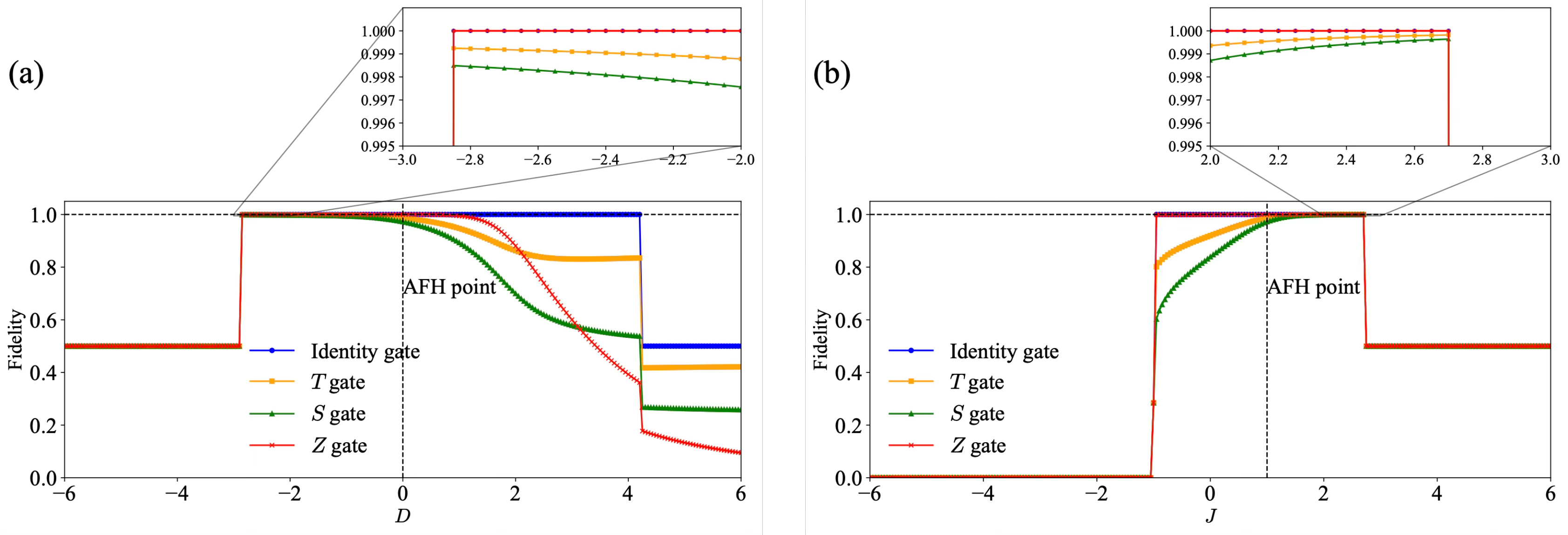}
\caption{(a) The gate fidelity of $R_z(\theta)$ for the ground state of the $XXZ$ model with the system sizes $L=12$ and $J=1,~D \in [-6,6]$. (b) The gate fidelity of $R_z(\theta)$ for the ground state of the $XXZ$ model with the system sizes $L=12$ and $J\in [-6,6],~D=0$. "AFH point" means the AFM Heisenberg point corresponding to $J=1,~D=0$. We calculate the gate fidelity of the identity, $T$-, $S$-, and $Z$ gates ($\theta = 0, ~\pi/4, ~\pi/2, ~\pi$). The ground state is obtained using DMRG  with a maximum bond dimension of 100.}
\label{fig:anis_AFHeis_RzGF}
\end{center}
\end{figure*}
ends is given by 
\begin{align}
&H_{XXZ}(J,D)=s^x_{\rm in}{S^x_1}+s^y_{\rm in}{S^y_1}+Js^z_{\rm in}{S^z_1}\notag\\*
&+\sum_{i=1}^{L-1}({S^x_i}{S^x_{i+1}}+{S^y_i}{S^y_{i+1}}+J{S^z_i}{S^z_{i+1}})\notag\\*
&+D\sum_{i=1}^L{(S^z_i)^2}+S^x_{L}s^x_{\rm out}+S^y_{L}s^y_{\rm out}+J
S^z_{L}s^z_{\rm out},
\label{eq:XXZ_hamiltonian_MBQC}
\end{align}
\begin{figure*}[htbp]
\begin{center}
\includegraphics[width=\linewidth]{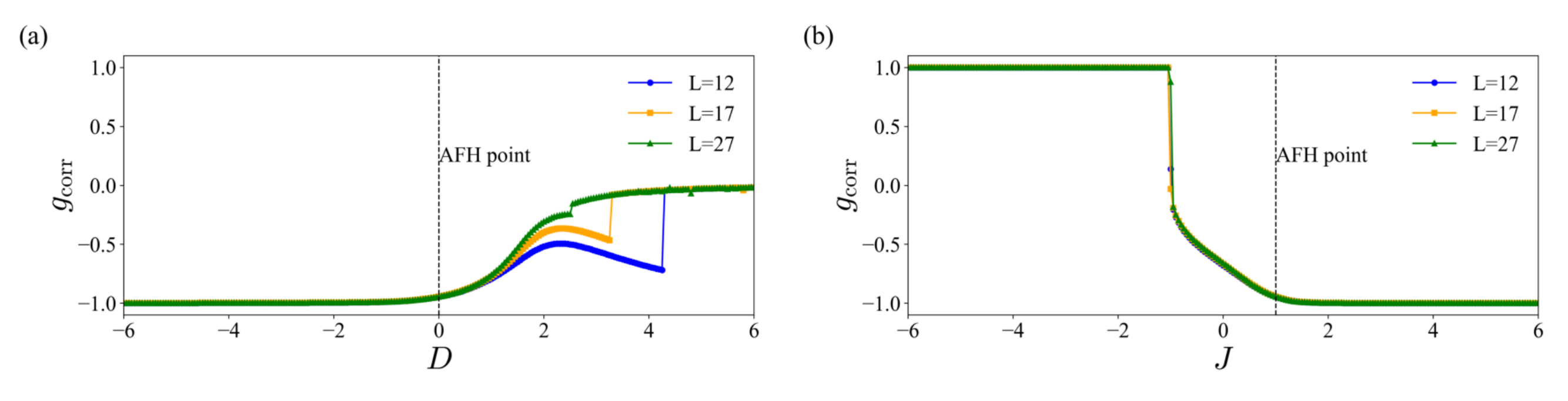}
\caption{The postmeasurement spin-spin correlation function defined in Eq.~(\ref{eq:g_GS}) for the ground state of Eq.~(\ref{eq:XXZ_hamiltonian_MBQC}) with (a) $D\in[-6,6]$ and (b) $J\in[-6,6]$. In both (a) and (b), results are  shown for $L=12,~17$, and $27$. The maximum bond dimension in the DMRG calculations sets as 100 for $L=12,~17$ and 200 for $L=27$.
}
\label{fig:XXZ_g_f}
\end{center}
\end{figure*}
where $J$ and $D$ control the Ising-like anisotropy and the single-ion anisotropy along the $z$-direction, respectively. The first and last three terms of the above Hamiltonian represent the coupling between spin-1 and spin-1/2 at the left and right ends. 

The standard spin-1 $XXZ$ model without spin1/2s at ends is known to exhibit a variety of ground state phases \cite{Chen2003,Shulz1986,Kennedy1992}, including FM, XY, Haldane, large-D, and AFM phases. In particular, the Haldane phase that emerges in this model is protected by $\mathbb{Z}_2 \times \mathbb{Z}_2$ symmetry and is thus characterized by a nonvanishing string order parameter \cite{Pollman2010,Tasaki1991}. 
Given that the Haldane phase appears in anisotropic regimes, it is of interest to study the effect of anisotropy on the computational power of the spin-1 $XXZ$ model as a resource state of MBQC.
\subsection{Gate fidelity of the $z$-rotation gate in the spin-1 $XXZ$ model}
\label{subsec:IVB}
We now focus on the MBQC performance of the spin-1 $XXZ$ model and numerically evaluate the gate fidelity, taking the resource state to be the ground state of Eq.~(\ref{eq:XXZ_hamiltonian_MBQC}). 

Figures~\hyperref[fig:XXZ_Id_S_GF]{5(a)} and \hyperref[fig:XXZ_Id_S_GF]{5(b)} show the gate fidelity of the identity and $S$ gate, respectively, evaluated with respect to the ground state of Eq.~(\ref{eq:XXZ_hamiltonian_MBQC}). 
As shown in Sec.~\ref{subsec:IIIB}, the gate fidelity for the identity gate is unity in the Haldane phase in Fig.~\hyperref[fig:XXZ_Id_S_GF]{5(a)}. In other phases, the gate fidelity is low.
Meanwhile, in Fig.~\hyperref[fig:XXZ_Id_S_GF]{5(b)}, the gate fidelity for the $S$ gate is not uniform within the Haldane phase. 
Remarkably, the gate fidelity for the $S$ gate is enhanced in the presence of anisotropy close to the phase boundary with the AFM phase.

In fact, the ground state in the Haldane phase near the AFM phase enables high-fidelity implementation of the $S$ gate as well as other rotation gates about the $z$ axis, as seen in Figs.~\hyperref[fig:anis_AFHeis_RzGF]{6(a)} and \hyperref[fig:anis_AFHeis_RzGF]{6(b)}, which
show the gate fidelity of the identity gate, $T$ gate, $S$ gate, and Pauli $Z$ gate with respect to the ground state of the Hamiltonian (\ref{eq:XXZ_hamiltonian_MBQC}) setting $J=1$ and $D=0$, respectively. In the presence of crystal field $D<0$ or Ising-like anisotropy $J>1$, the gate fidelity for the $T$ gate, $S$ gate and Pauli $Z$ gate increases as the ground state approaches the AFM phase, and reaches the maximum values around 0.99 when $D\sim-2.85$ and $J\sim2.70$. Thus, the AFM correlations in the Haldane phase enhance the computational power. However, when the anisotropy becomes stronger than these values, a phase transition from the Haldane phase to the AFM phase takes place, resulting in a sharp drop of the gate fidelity.
\subsection{Gate fidelity in the Haldane phase}
\label{subsec:IVC}
To understand the enhancement of gate fidelity in the presence of anisotropy, we analyze the gate fidelity of a general ground state in the Haldane phase, assuming only that it possesses the $\mathbb{Z}_2\times\mathbb{Z}_2$ symmetry. 
We now let $\ket{G} $ as a many-body state within the Haldane phase protected by $\mathbb{Z}_2\times\mathbb{Z}_2$ symmetry.
Taking advantage of the fact that $\ket{G}$ satisfies Eq.~(\ref{eq:Z2Z2symmetry}),
we can derive a general formula for the gate fidelity of the $R_z(\theta)$ gate as
\begin{equation}
F_{R_z(\theta)} 
=1
-\frac{\sin^2{\theta}}{2}\left(1+g_{\rm corr}\right) 
-\frac{(1-\cos{\theta})}{2}g_{\rm fail}.
\label{eq:Rz_GF_simplify}
\end{equation}
Here, $g_{\rm corr}$ denotes the postmeasurement spin-spin correlation function, defined as
\begin{equation}
g_{\rm corr} = \sum_{k=0}^{L} \bra{G} \left[
\left( \prod_{i=1}^{k}|z\rangle_i\langle{z}|\right)  Z_{\rm in}
S^z_{k+1}\right] \ket{G},
\label{eq:g_GS}
\end{equation}
where $k=0$ and $L+1$ denote the spin-1/2's at the left and right ends respectively ($S^z_{0}\equiv Z_{\rm in}$, $S^z_{L+1}\equiv Z_{\rm out}$) and the expression inside the square brackets is assumed to be the identity operator in the case of $k=0$. $g_{\rm corr}$ represents the correlation between $Z_{\rm in}$ and $S^z_{k+1}$ after the spins at the sites $i=1,\ldots,k$ are measured.
In Eq.~(\ref{eq:Rz_GF_simplify}), $g_{\rm fail}$ is given by
\begin{equation}
g_{\rm fail} = \bra{G}\left( \prod_{i=1}^{L}|z\rangle_i\langle{z}|\right) \ket{G}.
\label{eq:f_GS}
\end{equation}
It represents the probability for the measurements of all spins at $1\le i \le L$ to fail.
The detail of the derivation of Eq.~(\ref{eq:Rz_GF_simplify}) is provided in Appendix {\ref{sec:Appendix. B}}.

Figures~\hyperref[fig:XXZ_g_f]{7(a)} and \hyperref[fig:XXZ_g_f]{7(b)} show $g_{\rm corr}$ as functions of $D$ and $J$, with $J=1$ and $D=0$, respectively, for the ground state of the Hamiltonian in Eq.~(\ref{eq:XXZ_hamiltonian_MBQC}). In both figures, $g_{\rm corr}$ approaches $-1$ in the region $D<0$ and $J>0$ due to the strong AFM correlation near the AFM phase within the Haldane phase. 
Furthermore, since the AFM correlation favors the spin-1's to occupy the $|{\pm}\rangle$ states while suppressing the $|\tilde{0}\rangle$ state, the probability of obtaining the failure outcome $|z\rangle$ is reduced. Consequently, $g_{\rm fail}$ approaches $0$. 
Since achieving high gate fidelity requires $g_{\rm corr}\to-1$ and $g_{\rm fail}\to0$, the ground state in the Haldane phase near the AFM phase yields high gate fidelity, as shown in Figs.~~\ref{fig:XXZ_Id_S_GF} and \ref{fig:anis_AFHeis_RzGF}.   
However, introducing the uniaxial anisotropy along the $z$-axis negatively affects the implementation of rotation gates about the $x$ and $y$ axes, since a resource state polarized toward the $|\pm\rangle$ eigenstates in the $z$ direction suppresses computational capability along the $x$ and $y$ directions.
This limitation makes the realization of arbitrary single-qubit unitary gates particularly challenging. We overcome this obstacle in the next section by partitioning the chain and applying anisotropy in different directions.
\begin{figure}[t]
\begin{center}
\includegraphics[width=\columnwidth]{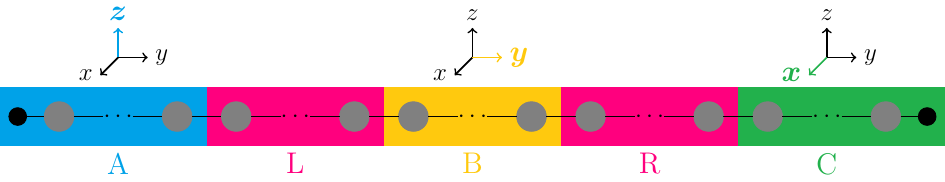}
\caption{Schematic illustration of division of the spin-1 chain. Anisotropy is introduced in the $z$, $y$, and $x$ directions in blocks A, B, and C, respectively, corresponding to the $R_z(\lambda)$, $R_y(\phi)$, and $R_x(\theta)$ gates. Blocks A, B, and C are constructed with $L/3$ spin-1 particles. And blocks ${\rm L}$ and ${\rm R}$ represent the isotropic blocks constructed with $N$ spin-1 particles.}  
\label{fig:3blocks_anisotropy_schematic}
\end{center}
\end{figure}
\begin{figure*}
\begin{center}
\includegraphics[width=\textwidth]{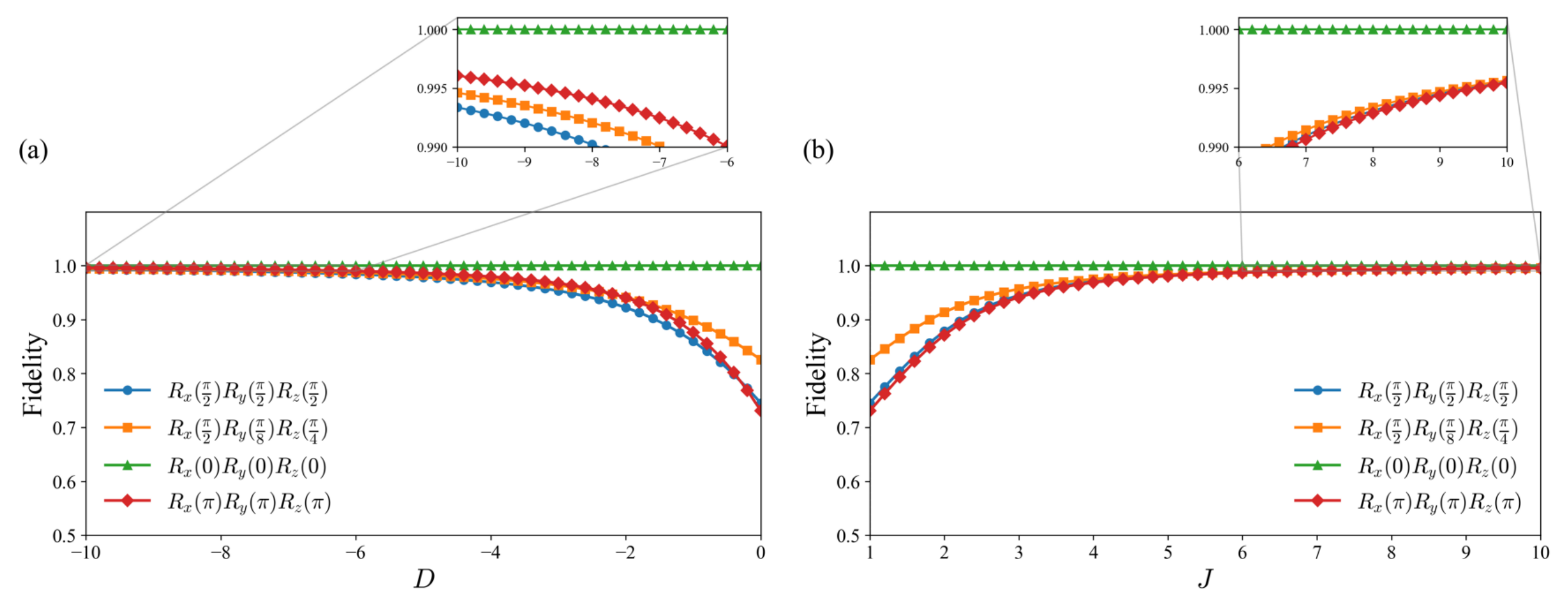}
\caption{The gate fidelity of the $R_x(\theta)R_y(\phi)R_z(\lambda)$ gate is evaluated for the parameter sets $(\theta,\phi,\lambda)=(\pi/2,\pi/2,\pi/2),~(\pi/2,\pi/8,\pi/4),~(0,0,0),~(\pi,\pi,\pi)$. (a) The gate fidelity for the ground state of Eq.~(\ref{eq:XXZ-XYX_XZZ_hamiltonian}) with the system sizes $L=9,~N=1$ and $J=1,~D\in[-10,0]$. (b) The gate fidelity for the ground state of Eq.~(\ref{eq:XXZ-XYX_XZZ_hamiltonian}) with the system sizes $L=9,~N=1$ and $J\in[1,10],D=0$. The ground state is obtained using DMRG  with a maximum bond dimension of 100.}
\label{fig:XXZ_XYX_XZZ_RxRyRzGF}
\end{center}
\end{figure*}

\section{Implementation of arbitrary single-qubit unitary gates}
\label{sec:V}
In this section, we demonstrate that arbitrary single-qubit unitary gates can be implemented via MBQC using the ground state of a spin-1 chain Hamiltonian as a resource state, by appropriately tuning the uniaxial anisotropy. 

An arbitrary single-qubit unitary gate $U$ can be decomposed into a product of three rotation gates about the $x$, $y$, and $z$ axes as \cite{Nielsen2011}
\begin{equation}
U = R_x(\theta)R_y(\phi)R_z(\lambda).
\label{eq:arbit_unitary}
\end{equation}
In order to implement this decomposed unitary gate in MBQC using a spin-1 chain as a resource, we divide an $L$-site spin-1 chain into three blocks, each consisting of $L/3$ sites, and label them as blocks A, B, and C. In addition, we insert $N$-site spin-1 junction blocks between the two adjacent blocks (see Fig.~\ref{fig:3blocks_anisotropy_schematic}). These spin-1 junction blocks are isotropic and are introduced to suppress competition between the anisotropies at the boundaries of each block. 
We implement $R_z(\lambda)$, $R_y(\phi)$, and $R_x(\theta)$ gates through measurements on blocks A, B, and C, respectively. In the junction blocks, the identity gate is performed to teleport the state from the previous block to the next one. The detailed measurement protocol is described in Appendix \ref{sec:Appendix. C1}.  

To implement the desired rotation gate in each block, anisotropy is applied along the corresponding direction in each block, as illustrated in Fig.~\ref{fig:3blocks_anisotropy_schematic}.
The Hamiltonian for implementation of the three rotation gates is given as
\begin{align}
&H_{XXZ-ZYZ-XYY}(J,D)=H_{\rm A}(J,D)+H_{\rm L}(J,D)\notag\\
&~~~~~~~~~~~~+H_{\rm B}(J,D) +H_{\rm R}(J,D)+H_{\rm C}(J,D),
\label{eq:XXZ-XYX_XZZ_hamiltonian}
\end{align}
\begin{align}
H_{\rm A}(J,D)=&{s^x_{\rm in}}{S^x_1}+s^y_{\rm in}{S^y_1}+Js^z_{\rm in}{S^z_1}\notag\\*
&+\sum_{i=1}^{L/3-1}({S^x_i}{S^x_{i+1}}+{S^y_i}{S^y_{i+1}}+J{S^z_i}{S^z_{i+1}})\notag\\*
&+D\sum_{i=1}^{L/3}{(S^z_i)^2},
\label{eq:H_A}
\end{align}
\begin{align}
H_{\rm B}(J,D)=&\sum_{i=L/3+N+1}^{2L/3+N-1}({S^x_i}{S^x_{i+1}}+J{S^y_i}{S^y_{i+1}}+{S^z_i}{S^z_{i+1}})\notag\\*
&+D{\sum_{i=L/3+N+1}^{2L/3+N}{(S^y_i)^2}},
\label{eq:H_B}
\end{align}
\begin{align}
H_{\rm C}(J,D)=&\sum_{i=2L/3+2N+1}^{L+2N-1}(J{S^x_i}{S^x_{i+1}}+{S^y_i}{S^y_{i+1}}+{S^z_i}{S^z_{i+1}})\notag\\*
&+D\sum_{i=2L/3+2N+1}^{L+2N}{(S^x_i)^2}+JS^x_{L+2N}s^x_{\rm out}\notag\\*
&+S^y_{L+2N}s^y_{\rm out}+S^z_{L+2N}s^z_{\rm out},
\label{eq:H_C}
\end{align}
\begin{align}
&H_{\rm L}(J,D)=\sum_{i=L/3}^{L/3+N}\bm{S}_i\cdot\bm{S}_{i+1},
\label{eq:H_L}
\end{align}
\begin{align}
&H_{\rm R}(J,D)=\sum_{i=2L/3+N}^{2L/3+2N}\bm{S}_i\cdot\bm{S}_{i+1},
\label{eq:H_R}
\end{align}
where $H_{\rm A}$, $H_{\rm B}$, and $H_{\rm C}$ correspond to blocks A, B, and C, respectively, and $H_{\rm L}$ and $H_{\rm R}$ correspond to junction blocks ${\rm L}$ and ${\rm R}$, respectively. $H_{\rm A}$, $H_{\rm B}$, and $H_{\rm C}$ include the Ising-like anisotropy $J$ and the single-ion anisotropy along the $x$, $y$, and $z$ directions, respectively. 
 
Figures~\hyperref[fig:XXZ_XYX_XZZ_RxRyRzGF]{9(a)} and \hyperref[fig:XXZ_XYX_XZZ_RxRyRzGF]{9(b)} show gate fidelity of the unitary gate in Eq.~(\ref{eq:arbit_unitary}) as functions $D$ and $J$, setting $J=1$ and $D=0$, respectively, where the ground state of the Hamiltonian in Eq.~(\ref{eq:XXZ-XYX_XZZ_hamiltonian}) serves as the resource state. The details of the calculation are given in Appendix C. These results are numerically obtained using the formula in Eq.~(\ref{eq:arbit_unitary_GF}).

In Fig.~\ref{fig:XXZ_XYX_XZZ_RxRyRzGF}, the gate fidelity is enhanced by applying a negative single-ion anisotropy ($D<0$) or an Ising-like anisotropy ($J>1$), and exceeds 0.99 when $D\lesssim-8.0$ or $J\gtrsim7.0$. In this parameter regime, the spin configurations in blocks A, B, and C are considered to be polarized toward the $\pm 1$ eigenstates of the $z$, $y$, and $x$ components, respectively, and to approach AFM ordering along the corresponding directions.

By introducing anisotropy along the direction corresponding to the rotation gate implemented in each of the blocks A, B, and C, MBQC based on the ground state of Eq.~(\ref{eq:XXZ-XYX_XZZ_hamiltonian}) can realize the unitary gate $U$ with higher fidelity than in the isotropic case with $J=1$ and $D=0$. In block A, as shown in the previous section, anisotropy along the $z$ axis ($J>1$ or $D<0$) enhances the gate fidelity of the rotation about the $z$ axis. Similarly, in blocks B and C, rotations about the $y$ and $x$ axes are achieved with high gate fidelity, since the failure states $|y\rangle$ and $|x\rangle$ are suppressed, leading to an enhancement of AFM correlation of the postmeasurement spins along the corresponding directions.
\section{CONCLUSION AND OUTLOOK}
\label{sec:VI}
In this work, we demonstrated that the ground state of a spin-1 $XXZ$ chain with uniaxial anisotropies, single-ion anisotropy $D$ and Ising-like anisotropy $J$, within the Haldane phase can serve as a resource state for MBQC implementing single-qubit gates. We evaluated the gate fidelity of both elementary rotation gates and general single-qubit unitary gates composed of rotations about the $x$, $y$, and $z$ axes, and found that the fidelity exceeds $0.99$ by appropriately tuning $D$ or $J$. Furthermore, we showed that the gate fidelity of a rotation gate, when the ground state lies in the Haldane phase, is characterized by the postmeasurement spin-spin correlation function and the failure probability. The high gate fidelity observed in the spin-1 $XXZ$ chain originates from the enhancement of AFM correlations near the AFM phase, which effectively suppresses the failure states.

Recent studies have proposed the realization of spin-1 chains with uniaxial anisotropies in cold atom systems \cite{Mogerle2025,Brechtelsbauer2025}. In particular, the spin-1 $XXZ$ Hamiltonian with such anisotropies has been predicted to emerge in systems of dysprosium atoms confined in an optical lattice \cite{Brechtelsbauer2025}. These platforms may provide an ideal testbed for our prediction that the ground state of the spin-1 $XXZ$ Hamiltonian can serve as a resource for MBQC, enabling high-fidelity implementation of single-qubit gates.
 
Our study is focused on performing the single-qubit unitary in spin-1 chains. Since the two-qubit gate is necessary to achieve universal quantum computation \cite{Deutsch1995,Deutsch1989}, an important next step would be to extend our study to  performing the two-qubit gate with high fidelity. 
\section*{ACKNOWLEDGEMENTS}
We thank D. Kagamihara and Y. Takeuchi for helpful discussion. ST is supported by JSPS KAKENHI Grant No. 25K07158.
\appendix
\section{THE GATE FIDELITY OF THE ROTATION GATE ALONG THE $z$ AXIS}
\label{sec:Appendix. A}
In this appendix, we present a brief derivation of Eq.~(\ref{eq:Rz_GF_expand}), following the treatment given in Ref.~\cite{masui2024}.
For the ideal state, we employ the following stabilizer representation \cite{Fujii2013,Gottesman1998}:
\begin{widetext}
\begin{equation}
|\psi_{R_z(\theta)}\rangle_{\rm in,out}\langle\psi_{R_z(\theta)}|=\left( \frac{I_{{\rm in}} I_{{\rm out}} + Z_{{\rm in}} Z_{{\rm out}}}{2} \right)\left( \frac{I_{{\rm in}} I_{{\rm out}} + X_{{\rm in}}  {R_{z,{\rm out}}(\theta)} 
X_{{\rm out}}[{R_{z,{\rm out}}(\theta)]^{\dagger}}}{2} \right).
\label{eq:stabilizer_form}
\end{equation}
The gate fidelity of $R_z(\theta)$, based on Eq.~(\ref{eq:GateFidelity}), can be written as
\begin{equation}
\begin{aligned}
F_{R_z(\theta)}=&\frac{1}{4} + \frac{1}{4} \Tr_{{\rm in}, {\rm out}}[\rho_{R_z(\theta)} Z_{{\rm in}} Z_{{\rm out}}]+ \frac{1}{4} \Tr_{{\rm in}, {\rm out}}\left\{\rho_{R_z(\theta)} X_{{\rm in}}  {R_{z,{\rm out}}(\theta)} X_{{\rm out}}[{R_{z,\rm out}(\theta)}]^{\dagger}\right\}\\
&+\frac{1}{4} \Tr_{{\rm in}, {\rm out}}\left\{\rho_{R_z(\theta)} X_{{\rm in}} Z_{{\rm in}}  {R_{z,{\rm out}}(\theta)} X_{{\rm out}} Z_{{\rm out}}[{R_{z,{\rm out}}(\theta)]^{\dagger}}\right\}.
\label{eq:Rz_GF}
\end{aligned}
\end{equation}\\
Using the expression in Eq.~(\ref{eq:state_aftermeasurement}), the second term in the above equation can be rewritten as follows:
\begin{equation}
\begin{aligned}
\Tr_{\rm in,out}\left[\rho_{R_z(\theta)}Z_{\rm in}Z_{\rm out}\right]&=\Tr_{\rm in,out}\left[\Tr_{\rm bulk}\left[\sum_{\bm{m}}B_{\rm out}^{(\bm{m})}{\mathcal{P}_{\bm{m}}}{\rho}{\mathcal{P}_{\bm{m}}}\left({B_{\rm out}^{(\bm{m})}}\right)^{\dag}\right]Z_{\rm in}Z_{\rm out}\right]\\
&=\Tr_{\rm all}\sum_{\bm{m}}\left[B_{\rm out}^{(\bm{m})}{\mathcal{P}_{\bm{m}}}{\rho}{\mathcal{P}_{\bm{m}}}\left({B_{\rm out}^{(\bm{m})}}\right)^{\dag}Z_{\rm in}Z_{\rm out}\right]\\
&=\Tr_{\rm all}\sum_{\bm{m}}\left[{\rho}Z_{\rm in}{\mathcal{P}_{\bm{m}}}\left({B_{\rm out}^{(\bm{m})}}\right)^{\dag}Z_{\rm out}B_{\rm out}^{(\bm{m})}{\mathcal{P}_{\bm{m}}}\right],
\label{eq:ZZterm}
\end{aligned}
\end{equation}
\\
where the trace  $\Tr_{\rm all}$ is taken over the entire system and $\rho=|G\rangle\langle{G}|$ denotes the resource state. Let $N_x(\bm{m})$ and $N_z(\bm{m})$ denote the number of $X$ and $Z$ by-products, respectively, associated with measurement outcomes $\bm{m}$. The by-product $B_{\rm out}^{(\bm{m})}$ can be represented as 
\begin{equation}
B_{\rm out}^{(\bm{m})}=X_{\rm out}^{N_x(\bm{m})+1}Z_{\rm out}^{N_z(\bm{m})+1}.
\label{eq:byprocuct_overall}
\end{equation}
Therefore, we have
\begin{equation}
\left({B_{\rm out}^{(\bm{m})}}\right)^{\dag}Z_{\rm out}B_{\rm out}^{(\bm{m})}{\mathcal{P}_{\bm{m}}}=(-1)^{N_x(\bm{m})+1}Z_{\rm out}{\mathcal{P}_{\bm{m}}}.
\label{eq:by-product_sign}
\end{equation}
To count the number of by-products, we use the following relation:
\begin{equation}
e^{-i\pi S^{\mu(\theta)}}|\nu(\theta)\rangle\langle{\nu(\theta)}|=
(-1)^{\delta_{\mu,\nu}}|\nu(\theta)\rangle\langle{\nu(\theta)}|.
\label{eq:Z2_proj}
\end{equation}
Then, Eq.~(\ref{eq:by-product_sign}) can be rewritten as follows:
\begin{equation}
\left({B_{\rm out}^{(\bm{m})}}\right)^{\dag}Z_{\rm out}B_{\rm out}^{(\bm{m})}{\mathcal{P}_{\bm{m}}}=-{\mathcal{P}_{\bm{m}}}\left(\prod_{i=1}^L e^{-i\pi S_i^z}\right)Z_{\rm out}.
\end{equation}
Therefore, we can reformulate Eq.~(\ref{eq:ZZterm}) as follows:
\begin{equation}
\begin{aligned}
\Tr_{\rm in,out}\left[\rho_{R_z(\theta)}Z_{\rm in}Z_{\rm out}\right]&=-\Tr_{\rm all}\left[{\rho}Z_{\rm in}\left(\prod_{i=1}^L e^{-i\pi S_i^z}\right)Z_{\rm out}\right]\\
&=\langle U^z \rangle.
\end{aligned}
\end{equation}
By applying the same transformation to the remaining terms in Eq.~(\ref{eq:Rz_GF}), we obtain the desired expression Eq.~(\ref{eq:Rz_GF_expand}).
\section{THE GATE FIDELITY OF THE ROTATION GATE ALONG THE $z$ AXIS FOR THE HALDANE PHASE}
\label{sec:Appendix. B}
In this appendix, we present the derivation of Eq.~(\ref{eq:Rz_GF_simplify}) which represents the computational power of the ground states in the Haldane phase. 
Equation~(\ref{eq:Rz_GF_eachterm}) can be rewritten as
\begin{equation}
\begin{aligned}
T_{\mu\nu}(\theta)= &\sum_{k=1}^L\sum_{{\bm{m}_k}} \bra{G}\left[\mathcal{P}_{{\bm{m}}_k}\left(\prod_{i=1}^k {e^{-i\theta S^z_i}}\right)\right]\left[{\sigma}^{\mu}_{{\rm in}}\left( \prod_{j=1}^{L}e^{-i\pi S^{\mu}_j} \right)
{{\sigma}^{\nu}_{{\rm out}}}\right]\ket{G}\\
&+\langle{G}|\left(\prod_{i=1}^L |z\rangle_i\langle{z}|\right)\left[{\sigma}^{\mu}_{{\rm in}}\left( \prod_{j=1}^{L}e^{-i\pi S^{\mu}_j} \right)
{{\sigma}^{\nu}_{{\rm out}}}\right]\ket{G}.
\label{eq:Rz_GF_eachterm_RW}
\end{aligned}
\end{equation}
In the above representation, we use the relation given by
\begin{equation}
\begin{aligned}
e^{-i\pi S^x_{\theta}}&=e^{\frac{-i\theta S^z}{2}}e^{-i\pi S^x}e^{\frac{i\theta S^z}{2}}\\
&=e^{-i\theta S^z}e^{-i\pi S^x}.
\label{eq:spin1_twistedZ2}
\end{aligned}
\end{equation}
By focusing on the site $k$ where the measurement succeeds and using Eq.~(\ref{eq:POVM_success}), we can expand Eq.~(\ref{eq:Rz_GF_eachterm_RW}) as follows:
\begin{align}
T_{\mu\nu}(\theta)=&\langle{G}|\sum_{k=1}^L \left[\left(\prod_{i=1}^{k-1}|z\rangle_i\langle{z}|\right)\left(|x(\theta)\rangle_k\langle{x(\theta)}|+|y(\theta)\rangle_k\langle{y(\theta)}|\right)e^{-i\theta S^z_k}\right]\notag\\*
&\times\left(\prod_{i=k+1}^{L}\sum_{\mu=x,y,z}|{\mu}\rangle_i\langle{\mu}|\right)
\left[\sigma^{\mu}_{\rm in}\left(\prod_{i=1}^L{e^{-i\pi S^{\mu}_i}}\right)\sigma^{\nu}_{\rm out}\right]|{G}\rangle+\langle{G}|\left(\prod_{i=1}^{L}{|z\rangle_i\langle{z}|}\right)\left[\sigma^{\mu}_{\rm in}\left(\prod_{i=1}^L{e^{-i\pi S^{\mu}_i}}\right)\sigma^{\nu}_{\rm out}\right]|{G}\rangle\notag\\*
=&\sum_{k=1}^L \langle{G}|\left[\left(\prod_{i=1}^{k-1}|z\rangle_i\langle{z}|\right)\times\left(|x(\theta)\rangle_k\langle{x(\theta)}|+|y(\theta)\rangle_k\langle{y(\theta)}|\right)e^{-i\theta S^z_k}\right]\left[\sigma^{\mu}_{\rm in}\left(\prod_{i=1}^L{e^{-i\pi S^{\mu}_i}}\right)\sigma^{\nu}_{\rm out}\right]|{G}\rangle\notag\\*
&+\langle{G}|\left(\prod_{i=1}^{L}{|z\rangle_i\langle{z}|}\right)\left[\sigma^{\mu}_{\rm in}\left(\prod_{i=1}^L{e^{-i\pi S^{\mu}_i}}\right)\sigma^{\nu}_{\rm out}\right]|{G}\rangle.
\label{eq:Rz_GF_eachterm_RW_RW}
\end{align}
In the above representation, the empty product is defined as the identity operator for $k=1,~L$. Using the relation
\begin{equation}
e^{-i\theta S^z}=\cos{\theta}\left(|x(\theta)\rangle\langle{x(\theta)}|+|y(\theta)\rangle\langle{y(\theta)}|\right)+|z\rangle\langle{z}|-i\sin{\theta}S^z,
\end{equation}
Eq.~(\ref{eq:Rz_GF_eachterm_RW_RW}) can be deformed as
\begin{equation}
\begin{aligned}
T_{\mu\nu}(\theta)=&\cos{\theta}\langle{G}|\left[\sigma^{\mu}_{\rm in}\left(\prod_{i=1}^L{e^{-i\pi S^{\mu}_i}}\right)\sigma^{\nu}_{\rm out}\right]|{G}\rangle+(1-\cos{\theta})\langle{G}|\left\{\left(\prod_{i=1}^{L}{|z\rangle_i\langle{z}|}\right)\left[\sigma^{\mu}_{\rm in}\left(\prod_{i=1}^L{e^{-i\pi S^{\mu}_i}}\right)\sigma^{\nu}_{\rm out}\right]\right\}|{G}\rangle\\
&-i\sin{\theta}\sum_{k=1}^L\langle{G}|\left\{\left(\prod_{i=1}^{k-1}|z\rangle_i\langle{z}|\right)S^z_k\left[\sigma^{\mu}_{\rm in}\left(\prod_{i=1}^L{e^{-i\pi S^{\mu}_i}}\right)\sigma^{\nu}_{\rm out}\right]\right\}|{G}\rangle.
\end{aligned}
\label{eq:Rz_GF_eachterm_RW_RW_RW}
\end{equation}
We now consider the case of $\mu=\nu$, in which the above representation can be rewritten as
\begin{equation}
T_{\mu\mu}(\theta)=-\cos{\theta}-(1-\cos{\theta})\langle{G}|\left(\prod_{i=1}^L|z\rangle_i\langle{z}|\right)|{G}\rangle+i\sin{\theta}\sum_{k=1}^L\langle{G}|\left[\left(\prod_{i=1}^{k-1}|z\rangle _i\langle{z}|\right)S^z_k\right]U^{\mu}|{G}\rangle,
\label{eq:Taa}
\end{equation}
where $U^{\mu}$ is defined in Eq.~(\ref{eq:Z2Z2symmetry}). 
For an arbitrary operator $V$ that anticommutes with $U$, the following holds \cite{yang2024}:
\begin{equation}
\langle{G}|VU^{\mu}|{G}\rangle=-\langle{G}|U^{\mu}V|{G}\rangle=-\langle{G}|V|{G}\rangle=\langle{G}|V|{G}\rangle=0.
\end{equation}
\end{widetext}
From the above relation, the third term in Eq.~(\ref{eq:Taa}) vanishes and Eq.~(\ref{eq:Taa}) can be simplified as 
\begin{equation}
T_{\mu\mu}(\theta)=-\cos{\theta}-(1-\cos{\theta})\langle{G}|\left(\prod_{i=1}^L|z\rangle_i\langle{z}|\right)|{G}\rangle,
\label{eq:Taa_simp}
\end{equation}
for $\mu=x,y$.
Similarly, we can deform Eq.~(\ref{eq:Rz_GF_eachterm_RW_RW_RW}) for $\mu\ne\nu$ as
\begin{equation}
\begin{aligned}
T_{xy}(\theta)&=-T_{yx}(\theta)\\
&=\sin{\theta}\sum_{k=1}^L\langle{G}|\left[\left(\prod_{i=1}^{k-1}|z\rangle_i\langle{z}|\right)Z_{\rm in}S^z_k\right]|{G}\rangle\\
&=\sin{\theta}\sum_{k=0}^L\langle{G}|\left[\left(\prod_{i=1}^{k}|z\rangle_i\langle{z}|\right)Z_{\rm in}S^z_{k+1}\right]|{G}\rangle\notag\\
&~~~~-\sin{\theta}\langle{G}|\left(\prod_{i=1}^{L}|z\rangle_i\langle{z}|\right)|{G}\rangle,
\end{aligned}
\label{eq:Txy_simp}
\end{equation}
where $k = 0$ and $L+1$ denote the “in" and “out"  sites, respectively ($S^z_
{L+1}\equiv Z_{\rm out}$), and the expression inside the square parentheses represents the identity for $k=0$.
By substituting Eqs.~(\ref{eq:Taa}) and (\ref{eq:Txy_simp}) into Eq.~(\ref{eq:Rz_GF_expand}), we obtain the desired expression Eq.~(\ref{eq:Rz_GF_simplify}).
\section{MBQC FOR ARBITRARY UNITARY GATES}
\label{sec:Appendix. C}
In this appendix, we describe the MBQC protocol for implementing the unitary gate in Eq.~(\ref{eq:arbit_unitary}) using a resource state divided into five blocks. Furthermore, we derive the formula for the gate fidelity of this unitary gate.
\subsection{MBQC protocol}
\label{sec:Appendix. C1}
We now consider a many-body state of $L+2N$ spin-1's and two spin-1/2's at left and right ends, which serves as a resource state implementing the unitary gate in Eq.~(\ref{eq:arbit_unitary}). Here, blocks A--C correspond to the spin-1 sites $1\le a\le L/3$, $L/3+N+1\le b\le 2L/3+N$ and $2L/3+2N+1\le c\le L+2N$, respectively.
The junction blocks L and R correspond to the spin-1 sites $L/3+1\le i_L\le L/3+N$ and $2L/3+N+1\le i_R\le 2L/3+2N$, respectively.

The measurement is performed from the first site in block A in a basis rotated around the $z$-axis. The measurements proceed sequentially until a successful rotation gate is achieved. Once a measurement succeeds, the remaining sites in that block are measured in the unrotated basis corresponding to the identity gate. If the entire measurement on block A fails, we proceed to block L and perform measurements to implement the identity gate. This procedure is then repeated for the subsequent blocks, proceeding in the order A-L-B-R-C. 

Here, we choose the measurement basis in blocks A--C as follows:
\begin{equation}
\begin{aligned}
&|\mu({\lambda}_{a})\rangle_{a}=e^{-i\frac{\lambda}{2}{S_{a}^{z}}}|\mu\rangle_{a},~({~ a\in[1,L/3]}),\\
&|\nu({\phi}_{b})\rangle_{b}=e^{-i\frac{\phi}{2}{S_{b}^{y}}}|\nu\rangle_{b},~({~b\in[L/3+N+1,2L/3+N]}),\\
&|\xi({\theta}_{c})\rangle_{c}=e^{-i\frac{\theta}{2}{S_{c}^{x}}}|\xi\rangle_{c},~({~c\in[2L/3+2N+1,L+2N]}).
\label{eq:Ugatebasis}
\end{aligned}
\end{equation}\\
where $\mu,\nu,\xi=x,y,z$. ${\theta}_c,\ {\phi}_b $ and ${\lambda}_a$ are defined as follows:
\begin{equation}
\begin{aligned}
&{\theta}_{c}=(-1)^{s_C}\theta,\\
&{\phi}_{b}=(-1)^{s_B+1}\phi,\\
&{\lambda}_{a}=(-1)^{s_A}\lambda.
\label{eq:Ugateangle}
\end{aligned}
\end{equation}
where $s_A$, $s_B$, and $s_C$ denote the numbers of by-product operators resulting from the previous measurements on sites $1$ to $a - 1$, $b - 1$, and $c - 1$, respectively.
Specifically, $s_A$, $s_B$, and $s_C$ count $X$, both $X$ and $Z$, and $Z$, respectively.

As an example, we consider the AKLT state as the resource state for the unitary gate in Eq.~(\ref{eq:arbit_unitary}). Let $\bm{A},~\bm{B},~\bm{C},~\bm{L}$, and $\bm{R}$ denote the measurement outcomes in blocks A, B, C, L, and R, respectively, which are given by 
\begin{equation}
\begin{aligned}
&\bm{A}= (\mu_{L/3},\ldots,\mu_{a'+1},\nu_{a'}({\lambda}_{a'}),z_{a'-1},\ldots,z_1),\\
&\bm{B}= (\mu_{2L/3+N},\ldots,\mu_{b'+1},\nu_{b'}({\phi}_{b'}),y_{b'-1},\ldots,y_{L/3+N+1}),\\
&\bm{C}= (\mu_{L+2N},\ldots,\mu_{c'+1},\nu_{c'}({\theta}_{c'}),x_{c'-1},\ldots,x_{2L/3+2N+1}),\\
&\bm{L}=(\mu_{L/3+N},\ldots,\mu_{L/3+1}),\\
&\bm{R}=(\mu_{2L/3+N+1},\ldots,\mu_{2L/3+2N}).
\label{eq:Ugateoutcome}
\end{aligned}
\end{equation}\\
Here, we define $\mu_j = x,y,z$ for $j\in[1,L+2N]$, and $a',~b'~{\rm and }~c',$ as the site indices at which the measurement first succeeds in blocks A, B and C, respectively. 
\onecolumngrid
\noindent Furthermore, we define $\nu_{a'}({\lambda}_{a'})$, $\nu_{b'}({\phi}_{b'})$, and $\nu_{c'}({\theta}_{c'})$ as
\begin{align}
&\nu_{a'}({\lambda}_{a'})=x_{a'}({\lambda}_{a'}),~y_{a'}({\lambda}_{a'}),\notag\\*
&\nu_{b'}({\phi}_{b'})=z_{b'}({\phi}_{b'}),~x_{b'}({\phi}_{b'}),\\*
&\nu_{c'}({\theta}_{c'})=y_{c'}({\theta}_{c'}),~z_{c'}({\theta}_{c'}).\notag
\end{align}
The state after the measurement, in the case that the above outcomes are obtained, can be written as
\begin{align}
&\left[\left(\prod_{i=L+2N}^{c'+1}\sigma^{{\mu}_i}_{\rm out}\right)R_{x,{\rm out}}({\theta}_{c'})X ^{\delta_{\nu_{c'}({\theta}_{c'}), y_{c'}({\theta}_{c'})}}_{\rm out}Z_{\rm out}X^{{c'}-2L/3-2N-1}_{\rm out}\right]\left(\prod_{i=2L/3+N+1}^{2L/3+2N}\sigma^{\mu_i}_{\rm out}\right)\notag\\
&\times\left[\left(\prod_{i=2L/3+N}^{b'+1}\sigma^{{\mu}_i}_{\rm out}\right)R_{y,{\rm out}}({\phi}_{b'})X ^{\delta_{\nu_{b'}({\phi}_{b'}), x_{b'}({\phi}_{b'})}}Z ^{\delta_{\nu_{b'}({\phi}_{b'}), z_{b'}({\phi}_{b'})}}(XZ)^{{b'}-L/3-N-1}_{\rm out}\right]\left(\prod_{i=L/3+1}^{L/3+N}\sigma^{\mu_i}_{\rm out}\right)\notag\\
&\times \left[\left(\prod_{i=L/3}^{a'+1}\sigma^{{\mu}_i}_{\rm out}\right)R_{z,{\rm out}}({\lambda}_{a'})X_{\rm out}Z ^{\delta_{\nu_{a'}({\lambda}_{a'}),y_{a'}({\lambda}_{a'})}}_{\rm out}Z^{{a'}-1}_{\rm out}\right]X_{\rm out}Z_{\rm out}|\phi^+\rangle_{\rm in,out}\notag\\
&=\left[\left(\prod_{i=L+2N}^{c'+1}\sigma^{{\mu}_i}_{\rm out}\right)X ^{\delta_{\nu_{c'}({\theta}_{c'}), y_{c'}({\theta}_{c'})}}_{\rm out}Z_{\rm out}X^{{c'}-2L/3-2N-1}_{\rm out}\right]\left(\prod_{i=2L/3+N+1}^{2L/3+2N}\sigma^{\mu_i}_{\rm out}\right)\\
&\times\left[\left(\prod_{i=2L/3+N}^{b'+1}\sigma^{{\mu}_i}_{\rm out}\right)X ^{\delta_{\nu_{b'}({\phi}_{b'}), x_{b'}({\phi}_{b'})}}_{\rm out}Z ^{\delta_{\nu_{b'}({\phi}_{b'}), z_{b'}({\phi}_{b'})}}_{\rm out}(XZ)^{{b'}-L/3-N-1}_{\rm out}\right]\left(\prod_{i=L/3+1}^{L/3+N}\sigma^{\mu_i}_{\rm out}\right)\notag\\
&\times \left[\left(\prod_{i=L/3}^{a'+1}\sigma^{{\mu}_i}_{\rm out}\right)X_{\rm out}Z ^{\delta_{\nu_{a'}({\lambda}_{a'}), y_{a'}({\lambda}_{a'})}}_{\rm out}Z^{{a'}-1}_{\rm out}\right]X_{\rm out}Z_{\rm out}R_{x,{\rm out}}(\theta)R_{y,\rm out}(\phi)R_{z,{\rm out}}(\lambda)|\phi^+\rangle_{\rm in,out}\notag,
\label{eq:stateAMUgate}
\end{align}
where $\mu = x,y,z$ and $\nu= x,y$. Since the by-products in the above equation can be removed via LOCC, we obtain the ideal state $U_{\rm out}|\phi^+\rangle_{\rm in,out}$.
\subsection{Gate fidelity}
\label{sec:Appendix. C2}
We now consider the gate fidelity of the above protocol. Similarly to Appendix \ref{sec:Appendix. A}, the ideal state can be written as follows:
\begin{equation}
|\psi_U\rangle_{\rm in,out}\langle\psi_U|=\left(\frac{I_{\rm in}I_{\rm out}+Z_{\rm in}U_{\rm out}Z_{\rm out}U^{\dag}_{\rm out}}{2}\right)\left(\frac{I_{\rm in}I_{\rm out}+X_{\rm in}U_{\rm out}X_{\rm out}U^{\dag}_{\rm out}}{2}\right).
\end{equation}
Then, the gate fidelity can be written as
\begin{equation}
F_{U}= \frac{1}{4} + \frac{1}{4} \Tr_{{\rm in}, {\rm out}}[\rho_U X_{{\rm in}} U_{\rm out}X_{{\rm out}}U^{\dag}_{\rm out}]- \frac{1}{4} \Tr_{{\rm in}, {\rm out}}[\rho_U Y_{{\rm in}}U_{\rm out}Y_{{\rm out}}U^{\dag}_{\rm out}]+ \frac{1}{4} \Tr_{{\rm in}, {\rm out}}[\rho_U Z_{{\rm in}}U_{\rm out}Z_{{\rm out}}U^{\dag}_{\rm out}].
\label{eq:U_GF}
\end{equation}
Using Eq.~(\ref{eq:arbit_unitary}), the above expression can be rewritten as follows:
\begin{equation}
F_{U}= \frac{1}{4} - \frac{1}{4} \Tr_{{\rm in}, {\rm out}}[\rho_U \tilde{X}_{{\rm in}} X_{{\rm out}}]- \frac{1}{4} \Tr_{{\rm in}, {\rm out}}[\rho_U \tilde{Y}_{{\rm in}}Y_{{\rm out}}]- \frac{1}{4} \Tr_{{\rm in}, {\rm out}}[\rho_U \tilde{Z}_{{\rm in}}Z_{{\rm out}}].
\label{eq:U_GF_RW}
\end{equation}
Here, we define $\tilde{X},\tilde{Y},\tilde{Z}$ as
\begin{equation}
\tilde{\sigma}^\mu=R_x(\theta)R_y(-\phi)R_z(\lambda)\sigma^{\mu}R_x(-\theta)R_y(\phi)R_z(-\lambda)
\end{equation}
for $\mu=x,y,z$. 

Furthermore, using the relations in Eqs.~(\ref{eq:ZZterm})--(\ref{eq:Z2_proj}) and (\ref{eq:spin1_twistedZ2}), each term in Eq.~(\ref{eq:U_GF_RW}) can be rewritten as follows:
\begin{equation}
\Tr_{\rm in,out}[\rho_U\tilde{\sigma^\mu}_{\rm in}\sigma^{\mu}_{\rm out}]=\Tr_{\rm all}\left[\rho Q^{\mu}_C Q^{\mu}_{R} Q^{\mu}_B Q^{\mu}_{L} Q^{\mu}_A\tilde{\sigma^\mu}_{\rm in}\sigma^{\mu}_{\rm out}\right],
\end{equation}
where $\rho$ is an arbitrary resource state. Here, $Q^{\mu}_A$, $Q^{\mu}_B$, $Q^{\mu}_C$, $Q^{\mu}_{L}$, and $Q^{\mu}_{R}$ are defined as 
\begin{align}
Q^{\mu}_A=\sum_{a'=1}^{L/3}\sum_{\bm{A}_{a'}}\left[\mathcal{P}_{{\bm{A}}_{a'}}\left(\prod_{i=1}^{a'}e^{-i{\theta}_{i}S^z_{i}}\right)\left(\prod_{j=1}^{L/3}e^{-i\pi S^{\mu}_{j}}\right)\right]+\left(\prod_{i=1}^{L/3}|z\rangle_i\langle{z}|e^{-i\pi S^\mu_{i}}\right),
\end{align}
\begin{align}
Q^{\mu}_B=\sum_{b'=L/3+N+1}^{2L/3+N}\sum_{\bm{B_{b'}}}\left[\mathcal{P}_{\bm{B_{b'}}}\left(\prod_{i=L/3+N+1}^{b'}e^{-i{\phi}_{i}S^y_{i}}\right)\left(\prod_{j=L/3+N+1}^{2L/3+N}e^{-i\pi S^{\mu}_{j}}\right)\right]+\left(\prod_{i=L/3+N+1}^{2L/3+N}|y\rangle_i\langle{y}|e^{-i\pi S^\mu_{i}}\right),
\end{align}
\begin{align}
Q^{\mu}_C=\sum_{c'=2L/3+2N+1}^{L+2N}\sum_{\bm{C}_{c'}}\left[\mathcal{P}_{\bm{C_{c'}}}\left(\prod_{i=2L/3+2N+1}^{c'}e^{-i{\phi}_{i}S^x_{i}}\right)\left(\prod_{j=2L/3+2N+1}^{L+2N}e^{-i\pi S^{\mu}_{j}}\right)\right]+\left(\prod_{i=2L/3+2N+1}^{L+2N}|x\rangle_i\langle{x}|e^{-i\pi S^\mu_{i}}\right),
\end{align}
\begin{align}
Q^{\mu}_{L}=\prod_{i=L/3+1}^{L/3+N}e^{-i\pi S^{\mu}_i},
\end{align}
\begin{align}
Q^{\mu}_{R}=\prod_{i=2L/3+N+1}^{2L/3+2N}e^{-i\pi S^{\mu}_i},
\end{align}
where ${{\bm{A}}_{a'}}$, ${{\bm{B}}_{b'}}$, and ${{\bm{C}}_{c'}}$ denote the measurement outcomes in blocks A, B, and C in the case where the measurements succeed at sites $a'$, $b'$, and $c'$, respectively.
Therefore, the gate fidelity can be represented as follows:
\begin{equation}
F_U=\frac{1}{4}-\frac{1}{4}\sum_{\mu=x,y,z}\Tr_{\rm all}\left[\rho Q^{\mu}_CQ^{\mu}_{R} Q^{\mu}_BQ^{\mu}_{L}Q^{\mu}_A\tilde{\sigma^\mu}_{\rm in}\sigma^{\mu}_{\rm out}\right],
\label{eq:arbit_unitary_GF}
\end{equation}
The results shown in Fig.~\ref{fig:XXZ_XYX_XZZ_RxRyRzGF} in Sec.~\ref{sec:V} are computed using this formula.
\twocolumngrid

\end{document}